\documentstyle[11pt,psfig]{l-aa}
\begin{document}
\thesaurus{
              06.        % A&A Section 06. The Sun
              (02.08.1;  % Hydrodynamics,
               02.19.1;  % Shock waves,
               02.23.1;  % Waves,
               06.03.1;  % Sun: chromosphere of,
               06.15.1)  % Sun: oscillations
                            }
\title{Acoustic wave propagation in the solar atmosphere}
\subtitle{IV.~Nonadiabatic wave excitation with frequency spectra}
\author{ J.~Theurer \inst{1}, P.~Ulmschneider \inst{1} and M.~Cuntz \inst{1,2}}
\institute{  Institut f\"ur Theoretische
             Astrophysik der Universit\"at
             Heidelberg, Tiergartenstr.~15,\newline
             D--69121 Heidelberg, Germany
             \and 
             Center for Space Plasma and Aeronomic Research (CSPAR),
             EB 136M, University of Alabama in Huntsville,\newline
             Huntsville, AL 35899, USA
} 
\date{ Received 15.~Oct.~1996; accepted 11.~Feb.~1997}
\maketitle
\begin{abstract}
We study the response of the solar atmosphere to 
excitations by large amplitude acoustic waves with
radiation damping now included. Monochromatic adiabatic
waves, due to unbalanced heating, generate continuously rising 
chromospheric temperature plateaus in which the low frequency 
resonances quickly die out. All non-adiabatic calculations 
lead to stable mean chromospheric temperature distributions
determined by shock dissipation and radiative cooling. 
For non-adiabatic monochromatic wave 
excitation, a critical frequency $\nu_{cr}\sim 1/25$~Hz is 
confirmed, which separates domains of different resonance 
behaviour. For waves of $\nu<\nu_{cr}$, the resonances decay, 
while for waves of $\nu>\nu_{cr}$ persistent resonance 
oscillations occur, which are perpetuated by shock merging. 
Excitation with acoustic frequency spectra produces distinct
dynamical mean chromosphere models where the detailed
temperature distributions depend on the shape of the assumed
spectra. The stochasticity of the spectra and the ongoing shock 
merging lead to a persistent resonance behaviour of the 
atmosphere. The acoustic spectra show a distinct shape evolution 
with height such that at great height a pure 3 min band becomes 
increasingly dominant. With our Eulerian code we did not 
find appreciable mass flows at the top boundary.
\keywords{ hydrodynamics -- shock waves -- waves -- 
                sun: chromosphere --
                sun: oscillations
               }
\end{abstract}
\section {Introduction}
A pronounced signal of velocity and temperature fluctuations 
in the solar chromosphere, seen in the Ca II H and K, H$\alpha$ 
and the Ca II infrared triplet lines, are the 3 min ($\nu=5.5$ 
mHz) oscillations. For detailed reviews of the 3 min oscillations see 
Deubner (1991), Fleck \& Schmitz (1991), Rutten \& Uitenbroek 
(1991), Rossi et al.~(1992), Carlsson \& Stein (1994) as well as 
Rutten (1995, 1996). 
 
Fleck \& Schmitz (1991) were the first to show that the 3~min 
oscillations can be explained as the basic cut-off frequency 
resonance of the chromosphere. This view has been confirmed both 
analytically and numerically by Kalkofen et al.~(1994) as well 
as by Sutmann \& Ulmschneider (1995a,b; henceforth called 
Papers I, II, respectively). For extended 
recent analytical work see also Sutmann et al.~(1997; Paper III).
The atmospheric resonance is due to a local 
oscillation of gas elements around their rest positions in 
hydrostatic equilibrium. 

Previously, Leibacher \& Stein (1981) in the wake of the very 
successful interpretation of the photospheric 5 min oscillations had 
explained the 3 min oscillations as a cavity mode. It is now 
universally accepted that the 5 min oscillations are acoustic 
waves trapped in a subphotospheric cavity between the 
temperature minimum and a refracting temperature rise in the 
solar interior. Leibacher \& Stein suspected that the 
chromospheric 3~min oscillation might be likewise explained by a 
chromospheric cavity, acting between the temperature minimum and 
the temperature rise of the chromosphere-corona transition 
layer. 

In our previous work, we have argued against a 
chromospheric cavity explanation, because the effect and 
viability of the 3 min oscillation as a local resonance both by 
the above mentioned analytical work and by time-dependent 
numerical calculations is now well established.
Nonetheless, it is highly 
likely that a chromospheric cavity resonance might occur {\it in 
addition} to the local resonance. Observationally, the $180^{\rm o}$ 
phase jump in the Na I D line points to such a cavity (Fleck \& 
Deubner 1989). Theoretical work by Fleck \& 
Schmitz (1991), Kalkofen et al.~(1994), Carlsson \& Stein 
(1994), Papers I \& II as well as by Cheng \& Yi (1996), however,
has only been one-dimensional (1-D) and often did not 
even include the existence of the transition layer in the 
analysis. It can easily be seen in 1-D simulations that acoustic 
waves will not be reflected by a steep temperature gradient. But 
this is only the case because of the very special geometry of purely 
vertical wave propagation. As soon as the wave propagation 
is inclined to the vertical, the temperature gradient acts 
to refract the acoustic wave field and the transition layer becomes a 
reflecting boundary. 

A third way to simulate the 3 min oscillation phenomenon is to 
avoid dwelling on mechanisms and simply excite the solar 
atmosphere with velocity fluctuations derived from Doppler shift 
observations from a low lying Fe I line (Lites et al.~1993).
Impressive work using this approach aimed
to explain the chromospheric bright point phenomenon has been 
carried out by Carlsson \& Stein (1994). Similar work using the 
Lites et al.~observations but employing a different numerical 
code has recently been reported by Cheng \& Yi (1996). These 
latter authors argue that the power in the 3 min band is already 
included in the observed signal and that high frequency acoustic 
wave power will not appreciably contribute to the 3 min band. 
This claim has to be taken with some caution as the code of 
these authors, different from that of Carlsson \& Stein, does 
not have an adaptive mesh capability and thus does not allow to 
treat shocks accurately.
Consequently, the code is not able to describe basic acoustic wave 
properties like the limiting shock strength behaviour and shock 
merging. Therefore, the code cannot correctly describe the transfer 
process of high frequency wave power into lower frequencies 
via shock formation and shock merging. 

It is precisely this process of shock merging and the resulting 
transfer of high frequency acoustic wave power into the low 
frequency range, which we want to study in greater detail in the 
present work. We continue here the work of Paper II. The reason 
for this interest is that in recent years it 
became clear that the acoustic energy generation in late-type 
stars is strongly tied to the Kolmogorov type turbulent cascade 
generated by the surface convection zones of these stars, which 
has an extensive high frequency tail (Musielak et al.~1994). It 
is thus of fundamental importance to understand the evolution of 
the acoustic wave spectrum with atmospheric height. It 
should be noted, however, that it is inherently difficult
to observe the 
high frequency acoustic wave power because the line contribution 
functions extend over several pressure scale heights (Judge 
1990) and one consequently has small modulation transfer functions
(e.g.~Deubner et al.~1988, Ulmschneider 1990).
Another important issue is the role of
the high and low acoustic wave frequencies in the chromospheric 
heating.
Finally there is the problem of how the observed 
prominent 3 min spectral component is to be related to a relatively 
flat acoustic spectrum produced by the Kolmogorov type turbulence 
in the convection zone obtained by Musielak et al.~(1994). 

It has been recognized for some time (Rammacher \& Ulmschneider 
1992, Fleck \& Schmitz 1993, Kalkofen et al.~1994) that for 
adiabatic waves the process of shock merging can greatly 
modify the atmospheric response and will eventually lead to a 
complete transformation of high frequency power into low frequency 
power. In Paper II, confirming work by Kalkofen et al.~ (1994), 
it was shown that the spectrum of adiabatic acoustic waves continuously 
changes towards longer wavelengths with increasing atmospheric height,
and that this is due to the process of shock merging 
by which short--period shock waves are 
converted to long--period waves. In Paper II
we found that for monochromatic wave excitation, a critical 
frequency $\nu_{cr}$ exists, below which the monochromatic waves stay 
dominant and above which monochromatic waves are completely 
obliterated by the generation of resonances through shock merging. 
Cuntz (1987), calculating time--dependent wave models for Arcturus,
also considered radiative damping.  He found that short--period
waves are converted to long--period ones via shock 
merging. In this work, however, fully developed shocks are 
already inserted at the inner boundary of the atmospheric shell 
and the development of photospheric acoustic waves into shocks 
is not calculated explicitly. 
%%
%%%%%%%%%%%%%%%%%%%%%%%%%%%%%%%%%%%%%%%%%%%%%%%%%%%%%%%%%%%%%%{1}%%%%%
%\begfigwid 6 cm
\begin{figure*}[t] 
\begin {minipage}[t]{0.95 \textwidth}
\psfig{figure=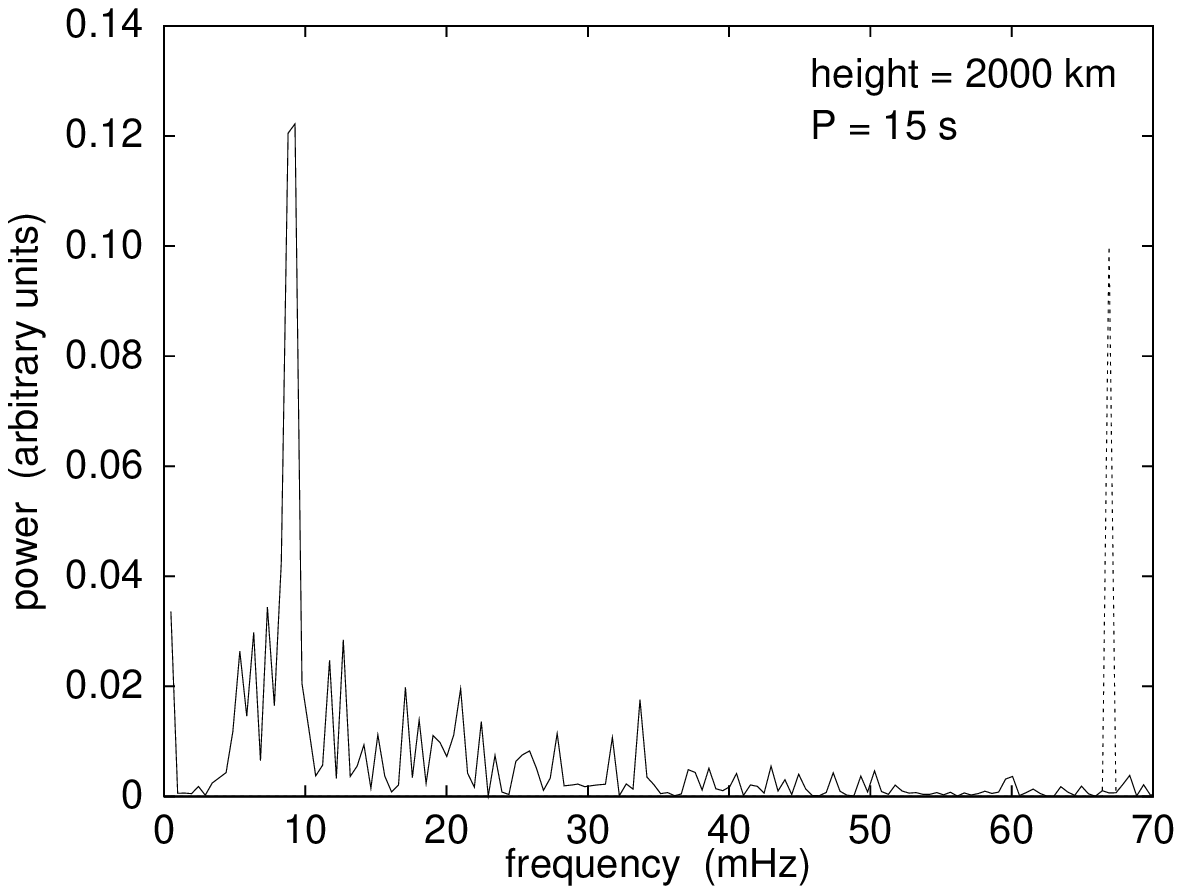,height=6.00cm,width=0.45 \textwidth}
\end{minipage}
\begin {minipage}[t]{0.95 \textwidth}
\psfig{figure=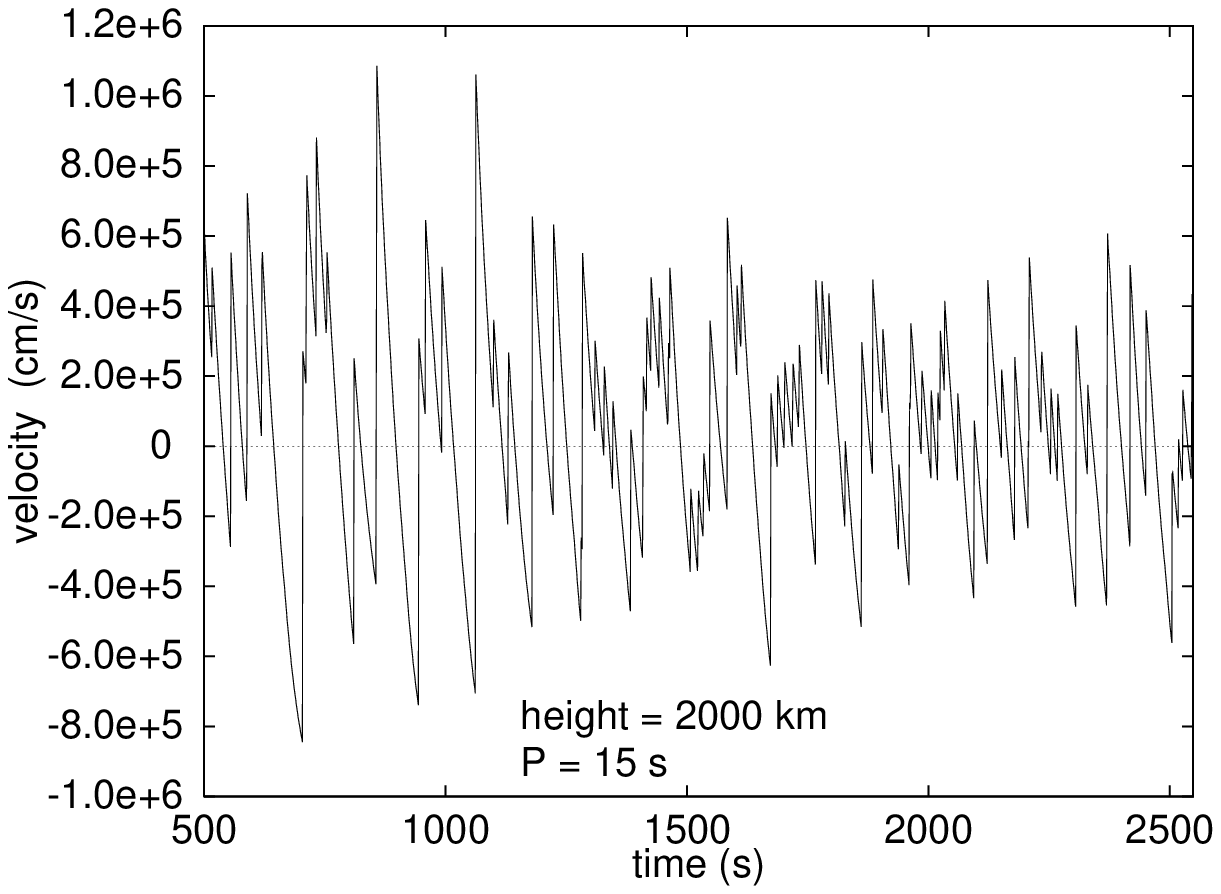,height=6.00cm,width=0.45 \textwidth}
\end{minipage}
\caption{
Power spectrum (left panel, drawn) and velocity (right panel) at height 
$z=2000$~km for an excitation with a monochromatic adiabatic 
wave of period $P=15$~s and flux
$F_{\rm A} = 1\cdot10^8$~erg~cm$^{-2}$~s$^{-1}.$
The power spectrum at height $z=0$~km, scaled by 0.1,
is shown dotted. The time span for the Fourier analysis is from 
500 to 2548 s, sampled by 1 s intervals.      
} 
\end {figure*} 
%%%%%%%%%%%%%%%%%%%%%%%%%%%%%%%%%%%%%%%%%%%%%%%%%%%%%%%%%%%%%%%%%%%%

%%%%%%%%%%%%%%%%%%%%%%%%%%%%%%%%%%%%%%%%%%%%%%%%%%%%%%{2}%%%%%%%%%%%%%%% 
%\begfigwid 6 cm
\begin{figure*}[t]
\begin{minipage}[t]{0.45 \textwidth}
\psfig{file=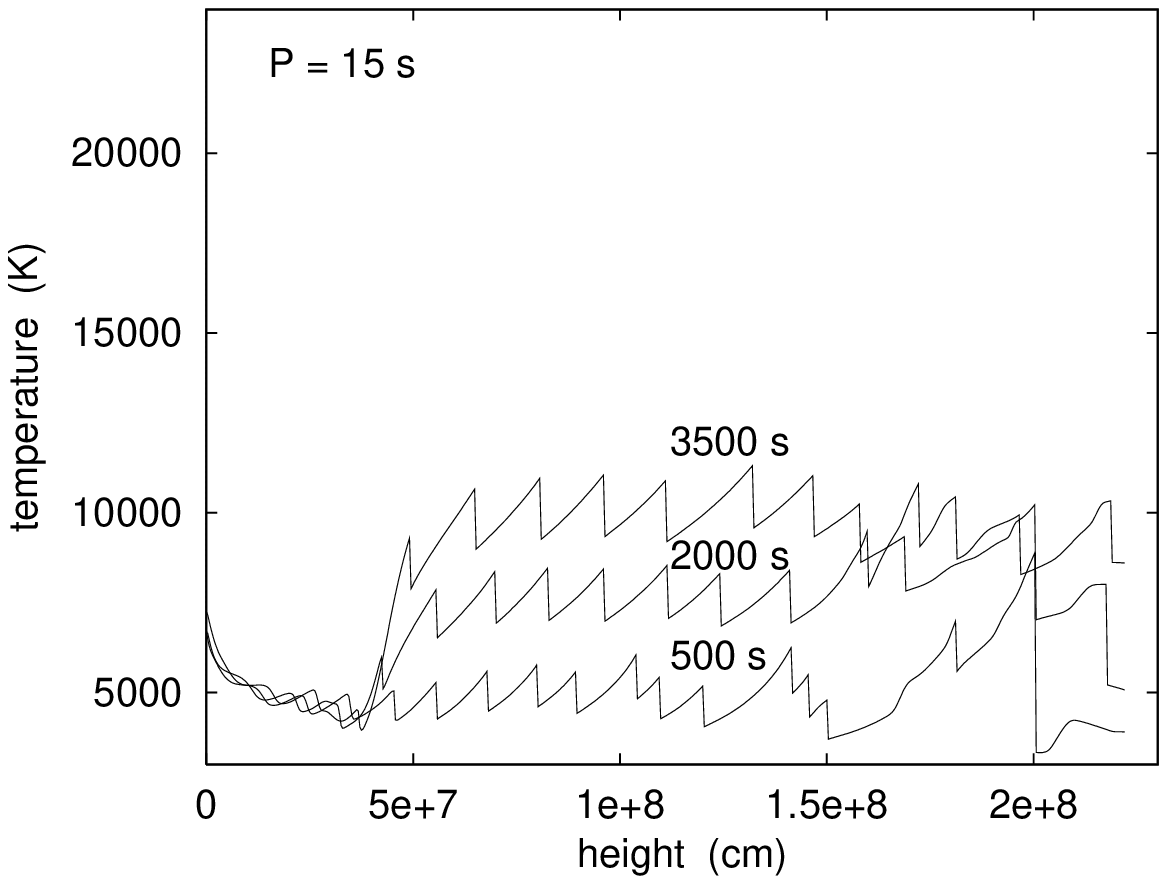,width=0.95 \textwidth,height=6.0cm}
\end{minipage}
\begin{minipage}[t]{0.45 \textwidth}
\psfig{file=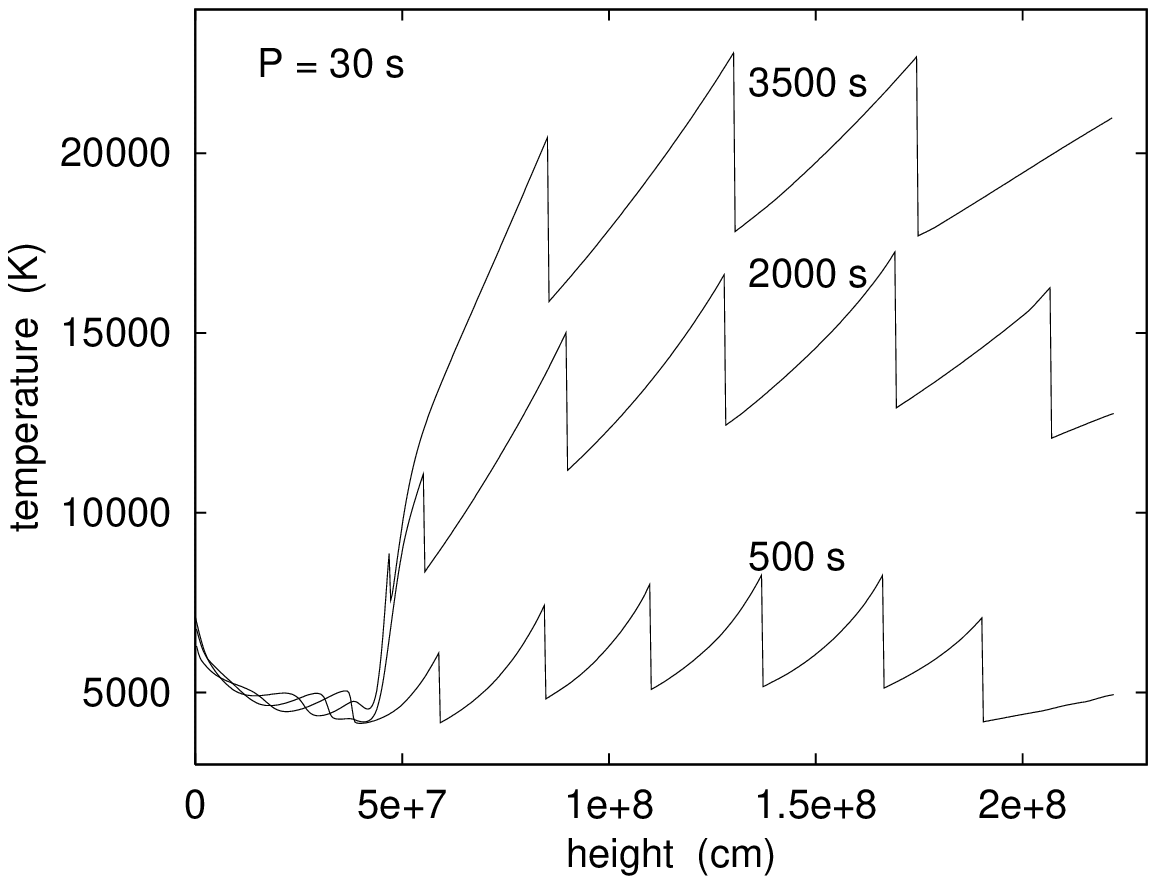,width=0.95 \textwidth,height=6.0cm}
\end{minipage}
\caption{
Temperatures as function of height at times $t=500, 
\ 2000,\ 3500$~s for a monochromatic adiabatic wave with an acoustic 
energy flux of
$F_{\rm A} = 1\cdot10^8$~erg~cm$^{-2}$~s$^{-1}$ and a period of
$P=15$~s (left panel) and $P=30$~s (right panel). 
}
\end{figure*}
%%%%%%%%%%%%%%%%%%%%%%%%%%%%%%%%%%%%%%%%%%%%%%%%%%%%%%%{2}%%%%%%%%%%%%%%%

In the present work we ask whether the theoretical results of
Paper I and II
persist when more realistic non-adiabatic wave calculations are 
performed and the time span of the calculations are extended.
We also want to increase the spectral resolution to make more
accurate predictions about the height evolution of the frequency 
spectra. Additionally,
we want to investigate the behaviour of monochromatic waves of 
much higher frequency. In Sect.~2 we briefly outline our
numerical methods. Section~3 presents the results and Sect.~4
gives our conclusions.
 
\section{Wave calculation method, Fourier analysis, 
atmosphere model} 
The method of wave computation has been described in Papers I and 
II and we refer to these papers for further details. 
The time--dependent 
hydrodynamic equations are solved using the method of 
characteristics. However, different to Papers I and II we now use 
an Eulerian description  
(Cuntz \& Ulmschneider 1988). We follow the development of the 
originally linear wave (introduced at the bottom of the 
atmosphere by a piston) to the point of shock formation and 
beyond. The shocks are treated as discontinuities and are 
allowed to grow to arbitrary strength. They are also permitted
to merge with each other.  We consider an atmospheric slab
with a height of 2220 km and use a total of 593 grid points
with a grid spacing of 3.75 km. In addition to the fixed number 
of regular grid points, there is an arbitrary number 
(typically 30) of shock points, which are allowed to move between the
regular grid points according to the speed of the shocks.  
Since we solve the hydrodynamic equation in Eulerian
form, our computational domain does not move with height and
artificial net outflows are avoided. 

Different from Papers I and II, we 
now consider radiation damping (see also Rammacher \& Ulmschneider 
1992, and Ulmschneider et al.~1992).  The radiative transfer and statistical 
equilibrium equations are solved for the H$^-$ continua and the 
Mg II $k$ and Ly $\alpha$ lines.  The lines are treated considering 
complete redistribution and using an operator splitting method 
(Buchholz et al.~1994). The radiative emission by the
Mg II $k$ line is scaled such as to
simulate a realistic total chromospheric radiation loss.
The usage of Fourier analysis and the evaluation of velocity power 
spectra is the same as described in Papers I and II, and we refer 
the reader to these papers for details. The
atmosphere models are discussed in Paper II. For our 
calculations we take a combined H$^-$, Mg II $k$ and Ly$\alpha$
radiative equilibrium model as initial atmosphere.
\section{Results} 
To investigate the properties of monochromatic waves we excited 
our H$^-$, Mg II $k$ and Ly$\alpha$ radiative equilibrium model by 
a piston at the bottom with waves of various periods and an 
acoustic flux of
$F_{\rm A} = 1\cdot10^8$~erg~cm$^{-2}$~s$^{-1}$.
Compared to Paper II we now calculate our models with a 2.5 times 
higher spatial resolution. As our time step is largely determined
by the Courant condition, this increases the temporal resolution by 
the same factor. 

\subsection{Monochromatic adiabatic wave excitation}
Because the wave calculation with $P=15$~s in Fig.~2 of Paper~II
was carried out with a low spatial resolution, we first 
calculate the adiabatic case again, using now about 
25 points per wavelength. Figure~1 shows the power spectra both at 
the bottom $z=0$~km (dotted) and at Euler height $z=2000$~km 
(left panel) together with the velocity at the latter height 
(right panel). The Fourier spectrum was computed for the time 
span 500 to 2548 s with the same sampling rate of 1 s as in 
the models of Paper II. Comparison with Fig.~2a of Paper II shows 
that, while that power spectrum is strongly peaked at about 8 
mHz, our present spectrum has extended power between 3 and 30 
mHz. This indicates that by decreasing the resolution 
a much narrower resonance oscillation is obtained, which can be 
explained by the fact that with higher spatial resolution the 
shock merging process is more individualistic, leading to a 
more pronounced low frequency spectrum. 

\subsection{Effect of the generated chromosphere model, adiabatic 
case}
In Paper II we did not discuss the important effect that, 
different from the linear waves in Paper I, non-linear waves 
will lead to significant heating, which considerably modifies the initial 
atmosphere and produces a dynamical chromosphere model. Figure~2 shows 
temperatures as function of height for adiabatic wave 
calculations with periods of $P=15$ and 30 s and an energy flux of
$F_{\rm A} = 1\cdot10^8$~erg~cm$^{-2}$~s$^{-1}$.
The waves are displayed at times $t=500$, 
2000 and 3500~s. It is seen that shock formation occurs in both models 
at heights near $z=400$~km. Within a short distance above 
this height, the waves attain a sawtooth profile with limiting 
shock strength and produce significant heating. 

%%%%%%%%%%%%%%%%%%%%%%%%%%%%%%%%%%%%%%%%%%%%%%%%%%{3}%%%%%%%%%%%%%%
%\begfig 6 cm
\begin{figure}[t]
\psfig{file=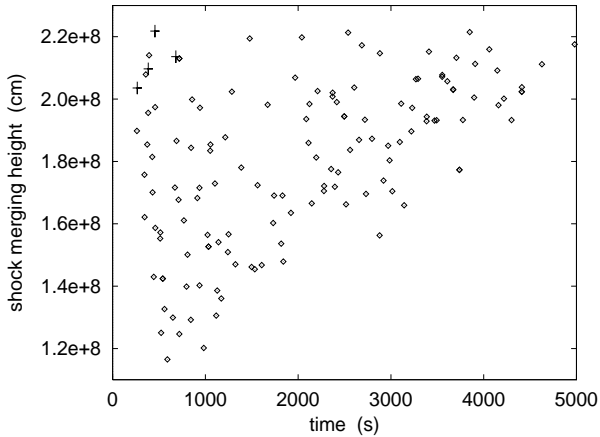,width=0.45 \textwidth,height=6.00cm}
\caption{
Shock merging heights as function of time
for monochromatic adiabatic waves with an acoustic energy flux of
$F_{\rm A} = 1\cdot10^8$~erg~cm$^{-2}$~s$^{-1}$ and a period of
$P=15$~s (diamonds) as well as 
$P=30$~s (plus signs). 
}
\end{figure}
%%%%%%%%%%%%%%%%%%%%%%%%%%%%%%%%%%%%%%%%%%%%%%%%%%{3}%%%%%%%%%%%%%%%
An interesting feature visible in Fig.~2 is that a chromospheric 
temperature plateau is formed which continuously rises as 
function of time. In the 15 s wave case the mean chromospheric 
temperature increases at a rate of 800 K every 500 s, from 
5000 K at $t=500$~s to 10000 K at $t=3500$~s. For the 30~s wave the 
increase is even faster with 2500 K per 500 s, from 6000 K at 
500 s to 21000 K at 3500 s. The reason for this continuously 
rising chromospheric temperature plateau is that despite of a 
perpetual heat input, the atmosphere has no 
way of cooling as the waves are treated adiabatically.
At a given height 
in the chromosphere each passing shock deposits an entropy jump 
$\Delta S$ per wave period which continuously increases the 
entropy. As for a given mass element the gas pressure stays 
roughly constant, the entropy increase leads directly to a 
temperature increase in the element.

Let us now discuss the effect of the atmosphere on the resonance 
oscillations. As discussed in Paper II and as seen in Fig.~1 
the initial switch-on effect dies out after about 1200 s 
at the height of 2000 km. Yet in Fig.~1 and in the non-adiabatic 
cases in Fig.~4 with periods $P=15$, 20, 30~s, there is a 
lot of low frequency power at times $t>1200$~s. This is due to 
the generation of resonances by shock merging events. Note that, 
as discussed in Paper II and by Rammacher \& Ulmschneider 
(1992), resonance oscillations are both the cause and the result of
shock merging events. This is because shocks, riding on the high 
temperature part of the low frequency resonant wave profile, catch up with
shocks propagating with lower speed on the low temperature part 
of that wave. The shock merging events kick on new transients, which 
leads to a prolongation of the resonances to times far beyond the 
decay time of the switch-on effect. 

We find that the atmospheric model strongly affects the shock
merging properties. The continuously increasing chromospheric
temperature plateau causes shock merging to occur at 
progressively greater heights. This is because (see Fig.~2) 
when two shocks of relatively similar strength move up a steep 
temperature rise, the second shock cannot easily catch up with 
the first as it moves with greater speed. Only when 
the temperature plateau is reached, can the second shock catch 
up. The persistently growing temperature of the plateau leads to 
an ever increasing height where the plateau starts (see Fig.~2) 
and thus where the merging events can take place. As the shock merging 
disturbance occurs in a progressively thinner atmosphere it 
generates weaker and weaker resonance oscillations, which in 
turn decreases the number of shock merging events with time.     

Figure~3 depicts the shock merging heights as function of time for 
two adiabatic wave computations with $P=15$~s (diamonds) and  
$P=30$~s (plus signs). As in the 30~s case the wavelength is 
much larger than in the 15~s case, it is seen that only 4 shock 
merging events happen in our computational domain, which extends to 
2220~km, and even then, shock merging occurs only in the initial switch-on 
phase at times $t < 700$~s. This allows the resonances to die 
out completely. In the 15 s case there are 108 shock merging events, 
but it is seen that with time these merging events occur at 
increasingly greater height. As discussed before, there will be a 
time when shock merging no longer occurs in the computational 
domain. Thus in the 15 s case the resonances will die out as 
well.
 
%%%%%%%%%%%%%%%%%%%%%%%%%%%%%%%%%%%%%%%%%%%%%%%%{4}%%%%%%%%%%%%%%%%%%%%%%
\begin{figure*}[t]
\begin{minipage}[t]{0.330 \textwidth}
\psfig{file=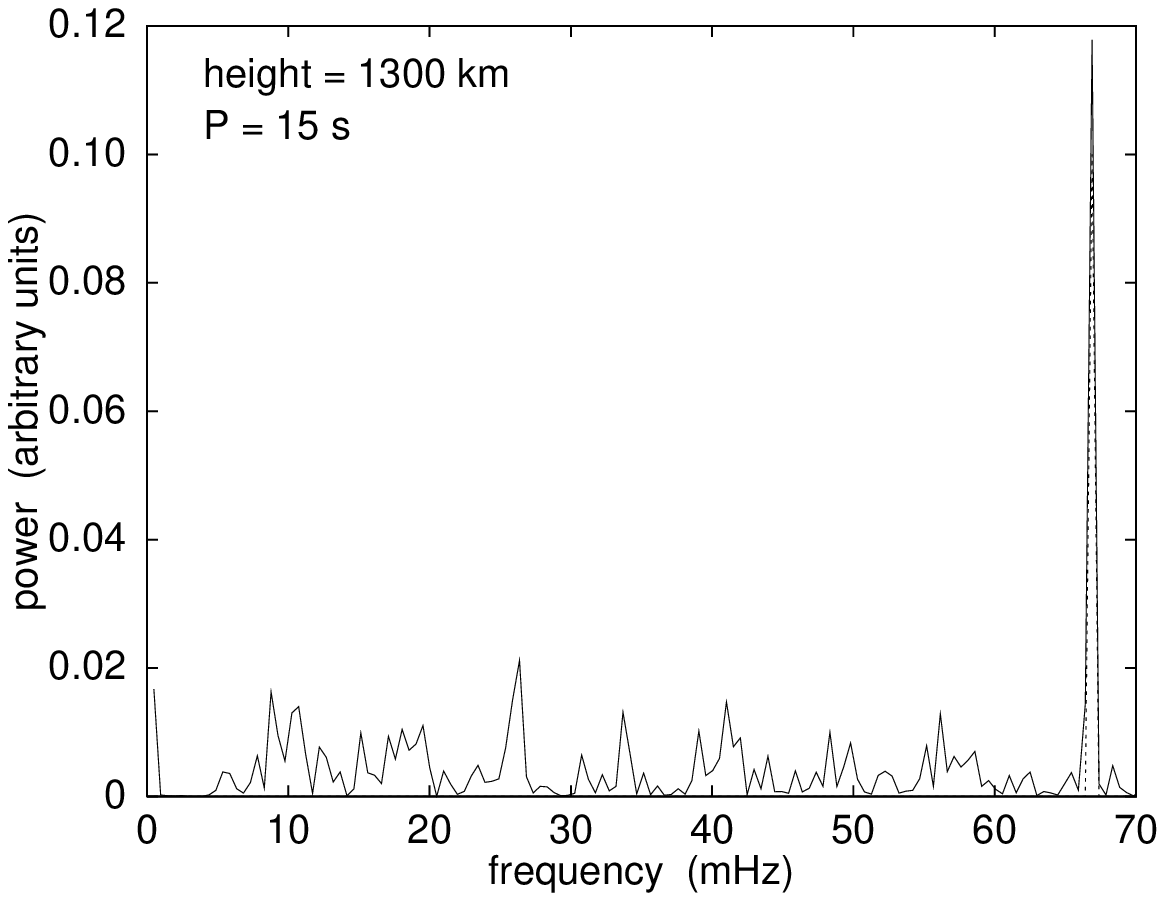,width=0.99 \textwidth,height=5.50cm}
\end{minipage}
\begin{minipage}[t]{0.330 \textwidth}
\psfig{file=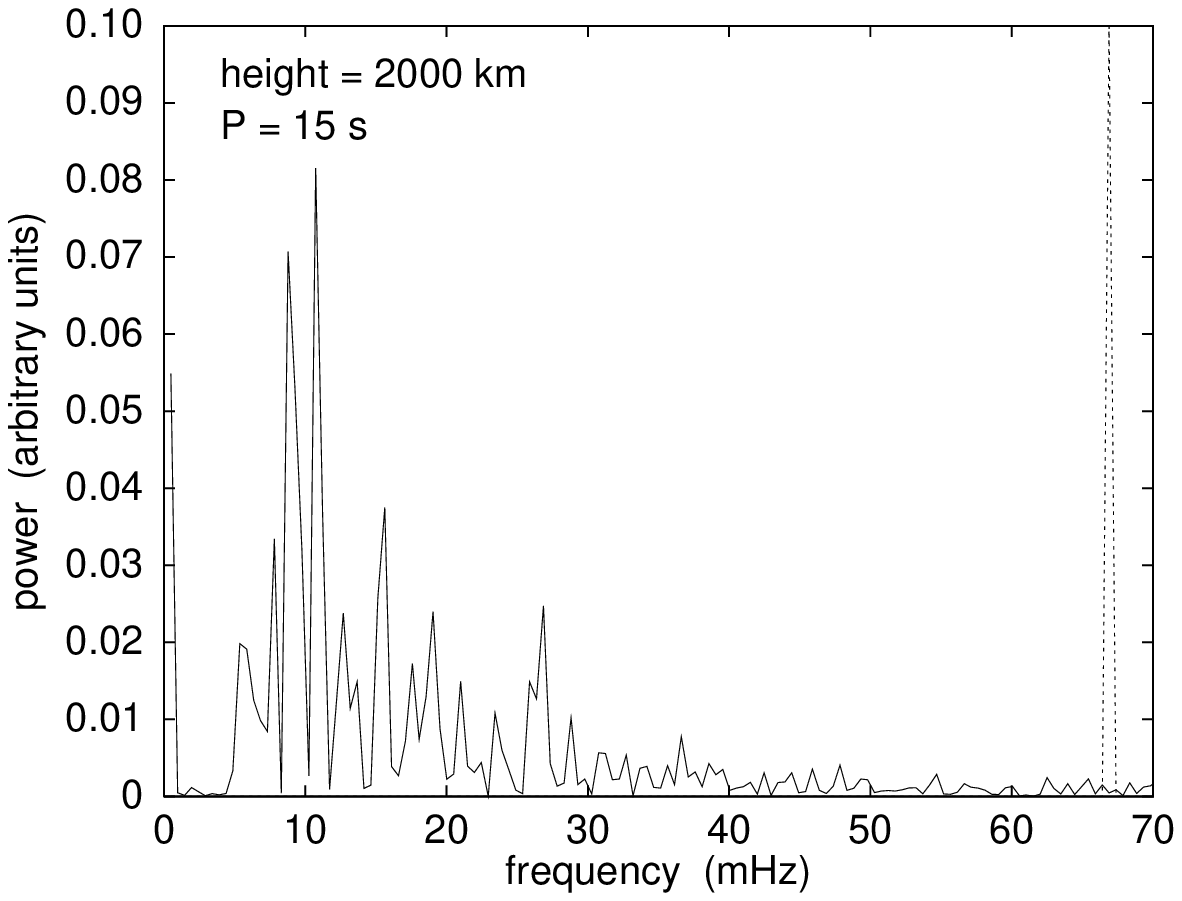,width=0.99 \textwidth,height=5.50cm}
\end{minipage}
\begin{minipage}[t]{0.330 \textwidth}
\psfig{file=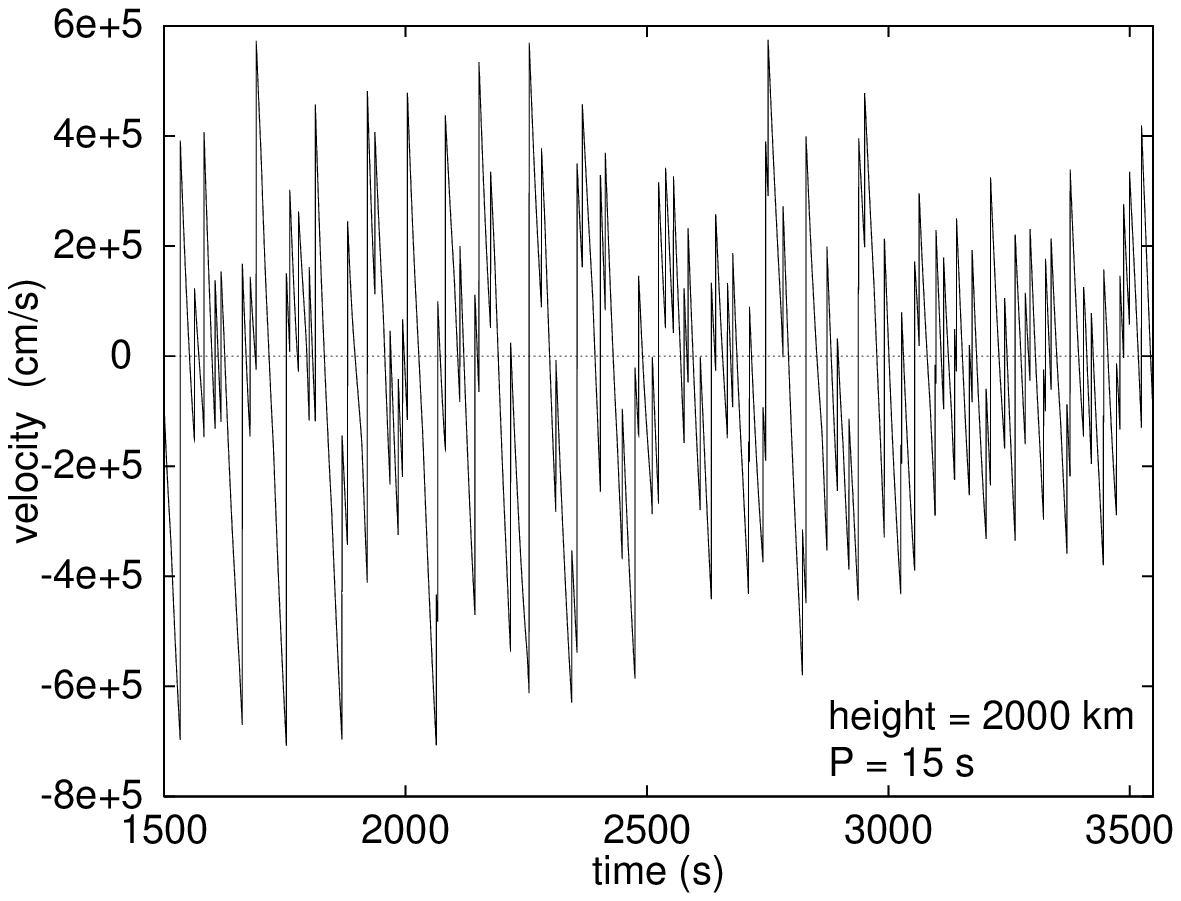,width=0.99 \textwidth,height=5.50cm}
\end{minipage}
%%%%%%%%%%%%%%%%%%%%%%%%%%%%%%%%%%%%%%%
\begin{minipage}[t]{0.330 \textwidth}
\psfig{file=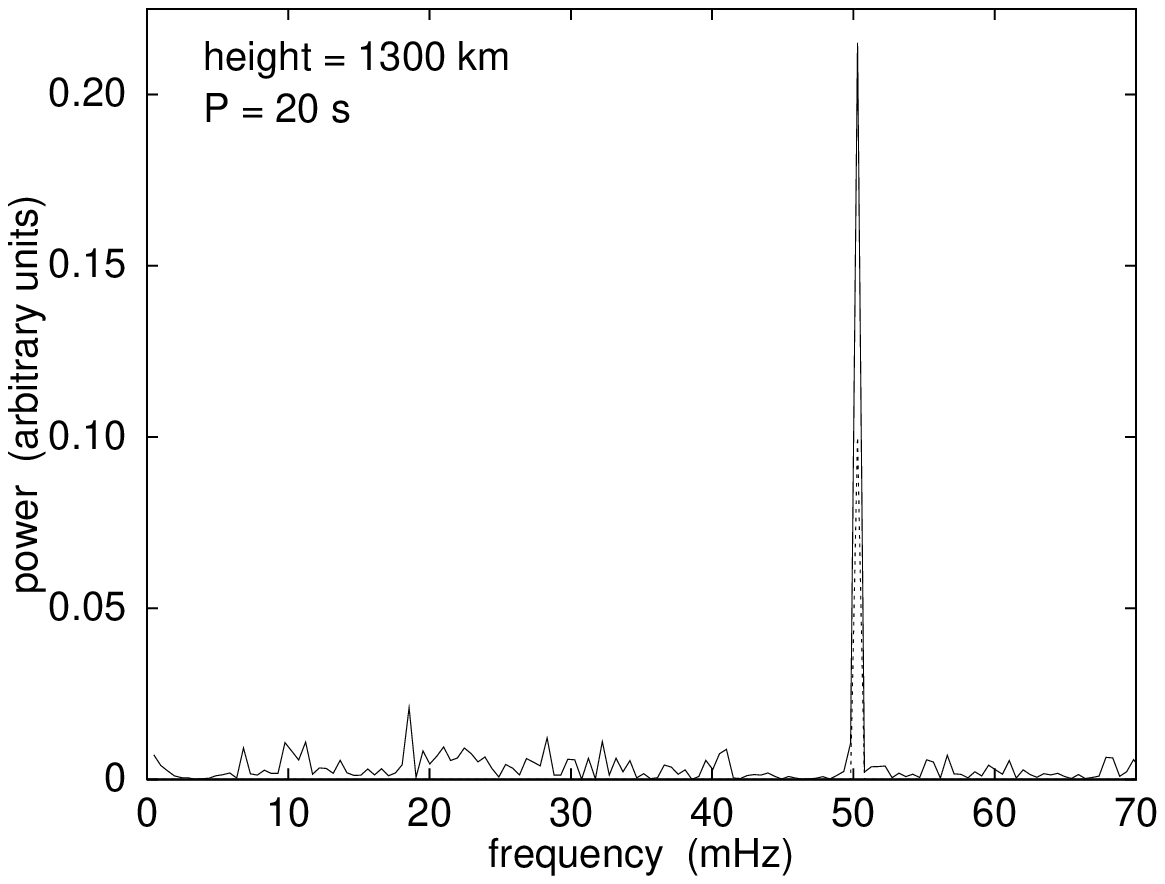,width=0.99 \textwidth,height=5.50cm}
\end{minipage}
\begin{minipage}[t]{0.330 \textwidth}
\psfig{file=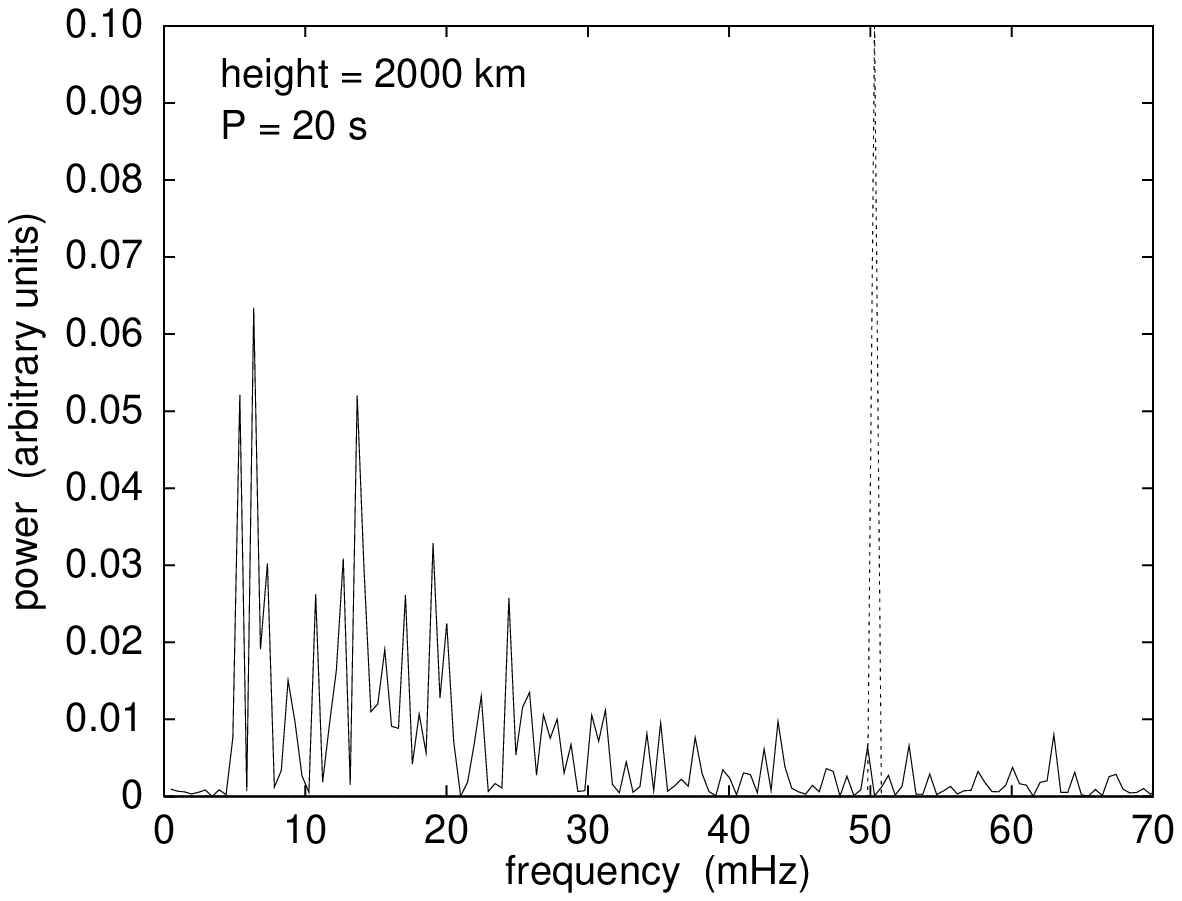,width=0.99 \textwidth,height=5.50cm}
\end{minipage}
\begin{minipage}[t]{0.330 \textwidth}
\psfig{file=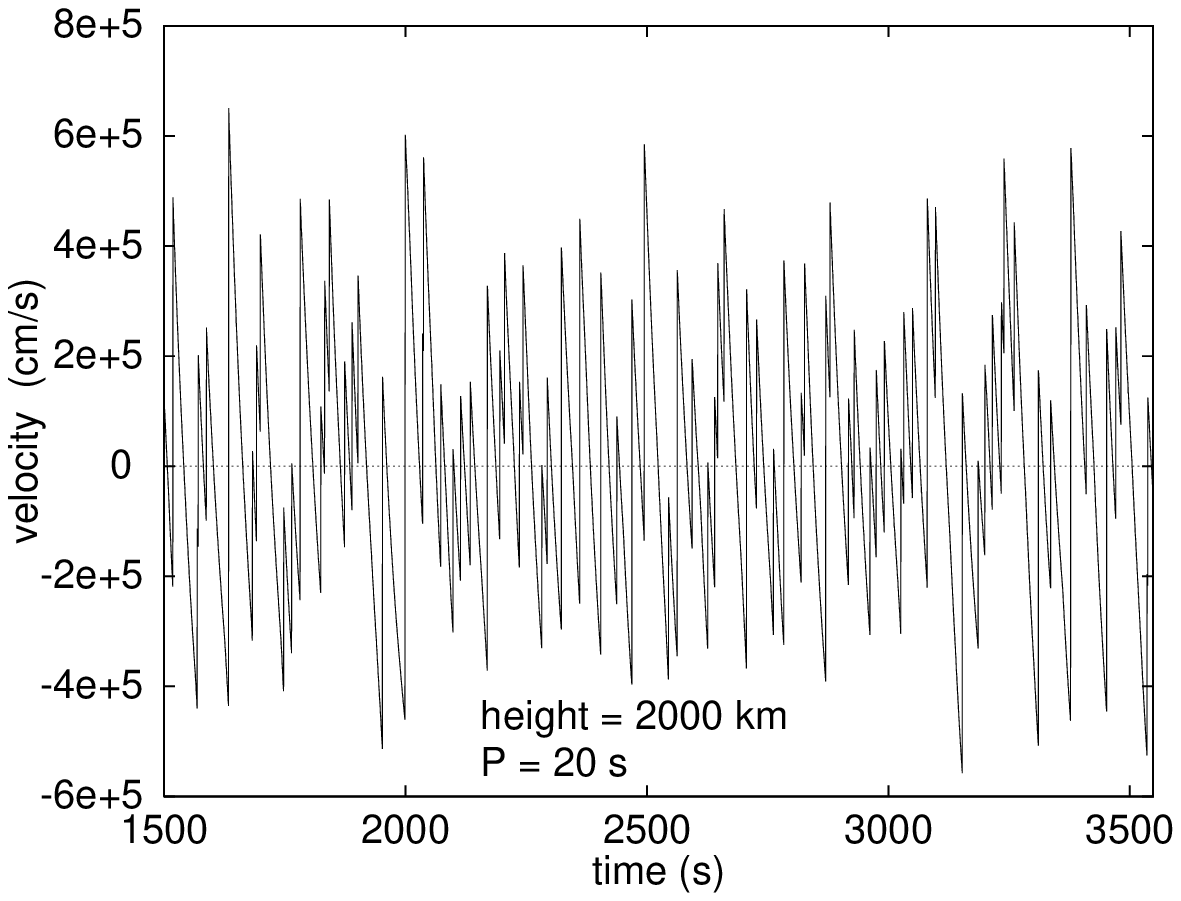,width=0.99 \textwidth,height=5.50cm}
\end{minipage}
%%%%%%%%%%%%%%%%%%%%%%%%%%%%%%%%%%%%%%%
\begin{minipage}[t]{0.330 \textwidth}
\psfig{file=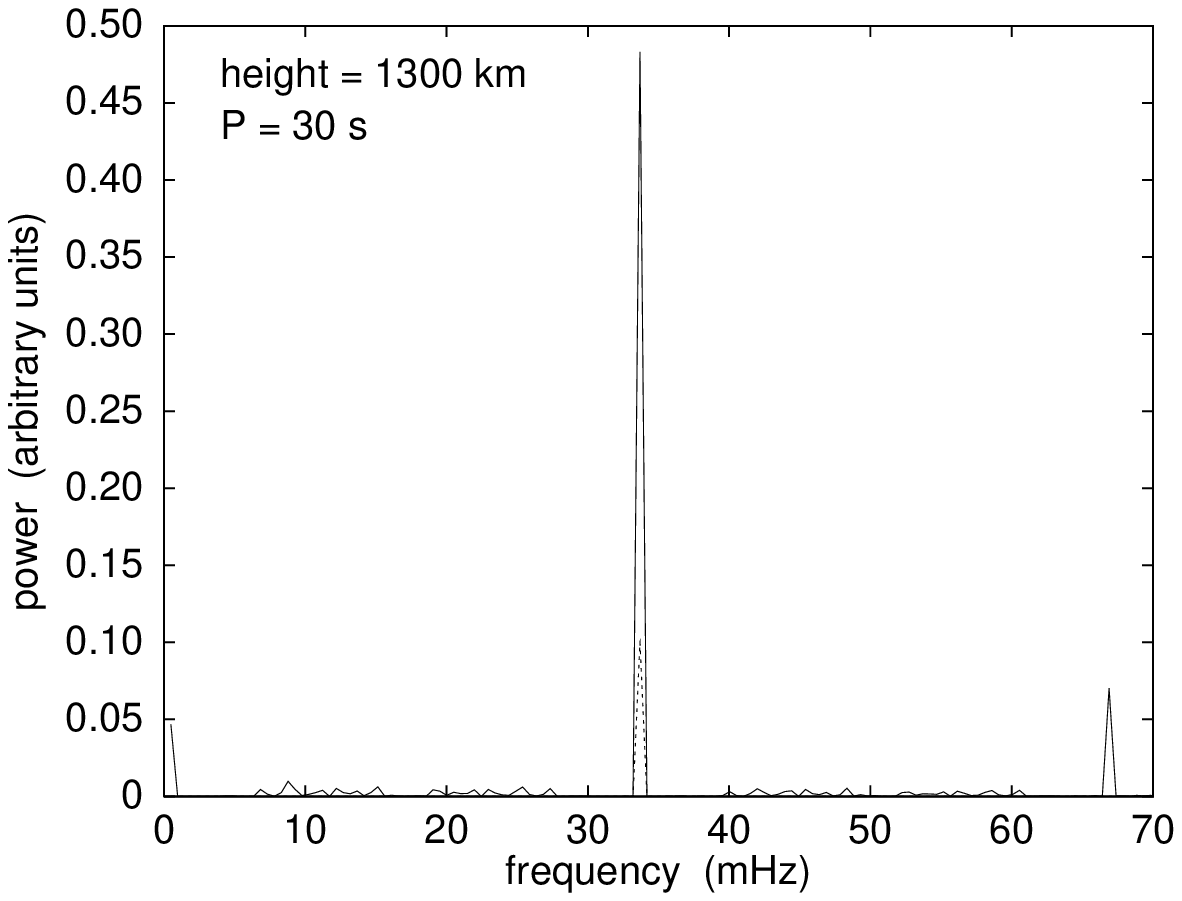,width=0.99 \textwidth,height=5.50cm}
\end{minipage}
\begin{minipage}[t]{0.330 \textwidth}
\psfig{file=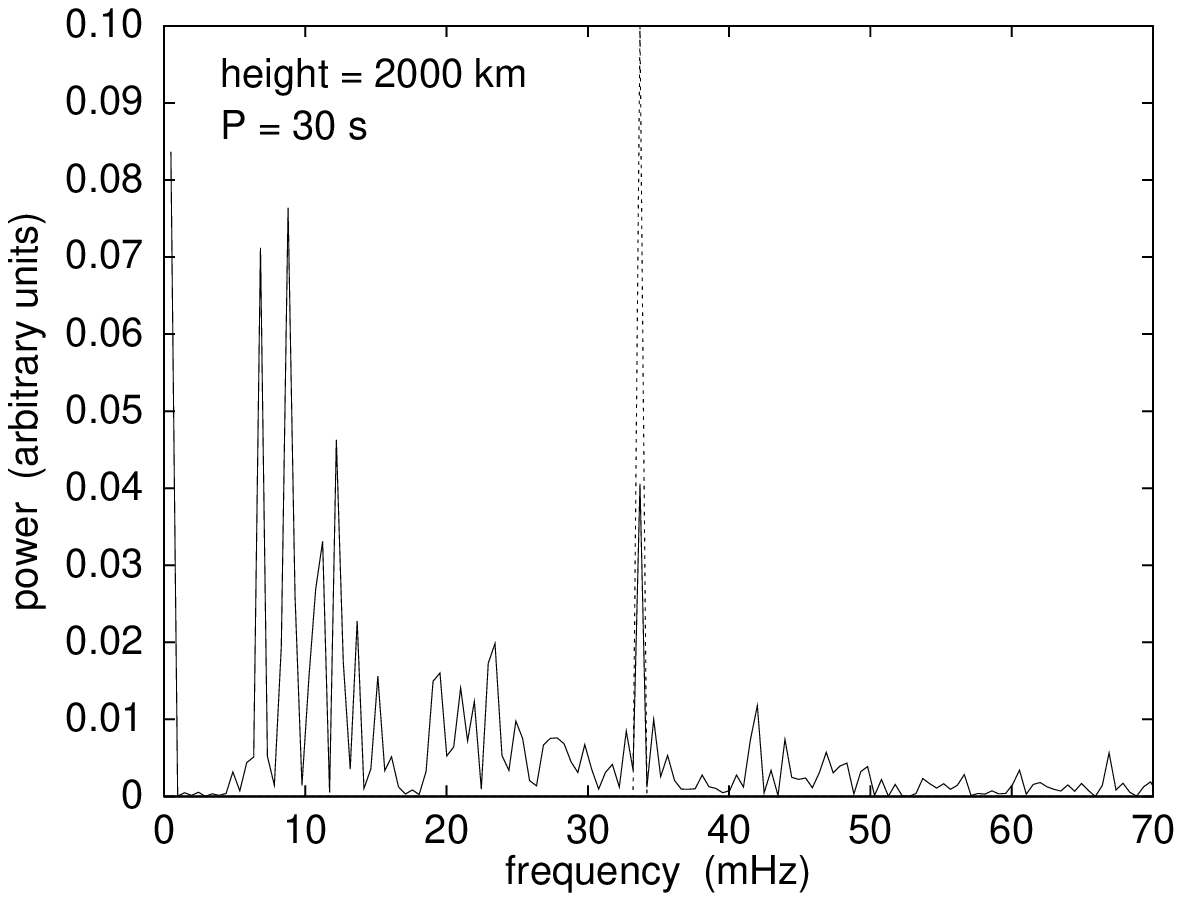,width=0.99 \textwidth,height=5.50cm}
\end{minipage}
\begin{minipage}[t]{0.330 \textwidth}
\psfig{file=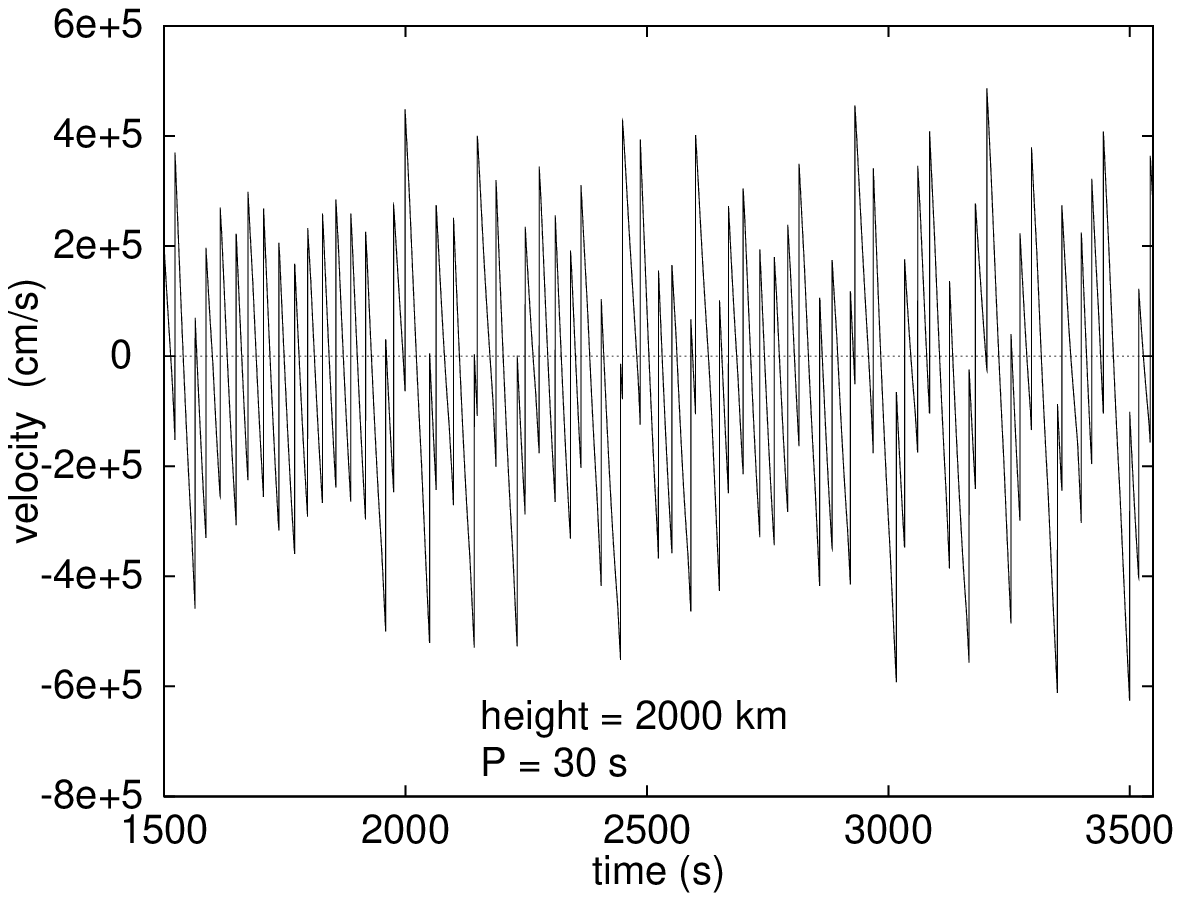,width=0.99 \textwidth,height=5.50cm}
\end{minipage}
\caption{
Power spectra (left and middle columns) and velocities at height 
$z=2000$~km (right column) for an excitation with monochromatic 
waves of period $P=15$, 20, 30~s (from top to bottom) and flux
$F_{\rm A} = 1\cdot10^8$~erg~cm$^{-2}$~s$^{-1}$,
for non-adiabatic waves. The power spectra 
at height $z=0$~km, scaled by 0.1,
are shown dotted, those at height $z=2000$~km 
drawn.  The velocity is sampled from 1500 to 3548 s in 1 s 
intervals. 
}
\end{figure*}
%%%%%%%%%%%%%%%%%%%%%%%%%%%%%%%%%%%%%%%%%%%%%%%%%%%%%%%%%%%%%%%%%%%%%%%
%

In summary we conclude that for adiabatic monochromatic waves, smaller wave
periods lead to more shock merging events which generate resonance 
oscillations. But the unrestricted heating in the adiabatic case 
modifies the atmospheric structure in such a way that 
eventually, at a given height, further shock merging is
prevented and the resonance oscillations die out. Thus for 
adiabatic non-linear monochromatic wave excitations, regardless 
of the wave period, 
the resonances will eventually die out and the 
forced oscillations will prevail. 
\subsection{Monochromatic non-adiabatic wave excitation} 
We now consider wave calculations with radiation damping. Figure~4 
(left two columns) depicts the power spectra at $z=0$, 1300 
and 2000~km for waves of period $P=15$, 20 and 30~s. These 
spectra differ in two important ways from those of Fig.~2 of 
Paper II. They are now computed allowing for radiation damping 
and the time span of the Fourier analysis is now from 1500 to 
3548 s instead of from 500 to 2048 s. The velocity amplitudes of 
the waves, displayed in the right column of Fig.~4 as well as 
in the corresponding Fig.~2 of Paper II, show that 
resonance oscillations occur well past the decay time of the 
initial switch-on effect. 

%%%%%%%%%%%%%%%%%%%%%%%%%%%%%%%%%%%%%%%%%%%%%{5}%%%%%%%%%%%%%%%%%%%%
\begin{figure*}[tb]
\begin{minipage}[t]{0.45 \textwidth}
\psfig{file=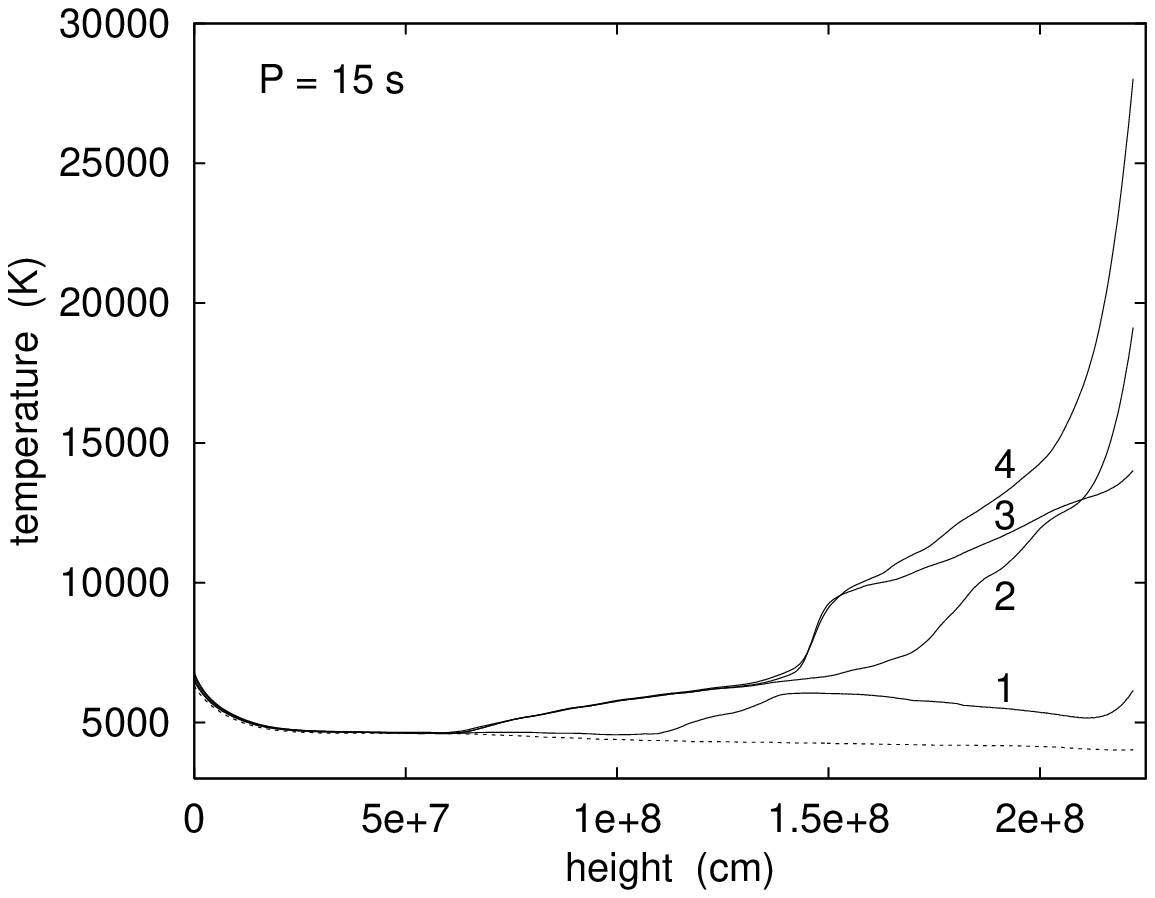,width=0.95 \textwidth,height=6.00cm}
\end{minipage}
\begin{minipage}[t]{0.45 \textwidth}
\psfig{file=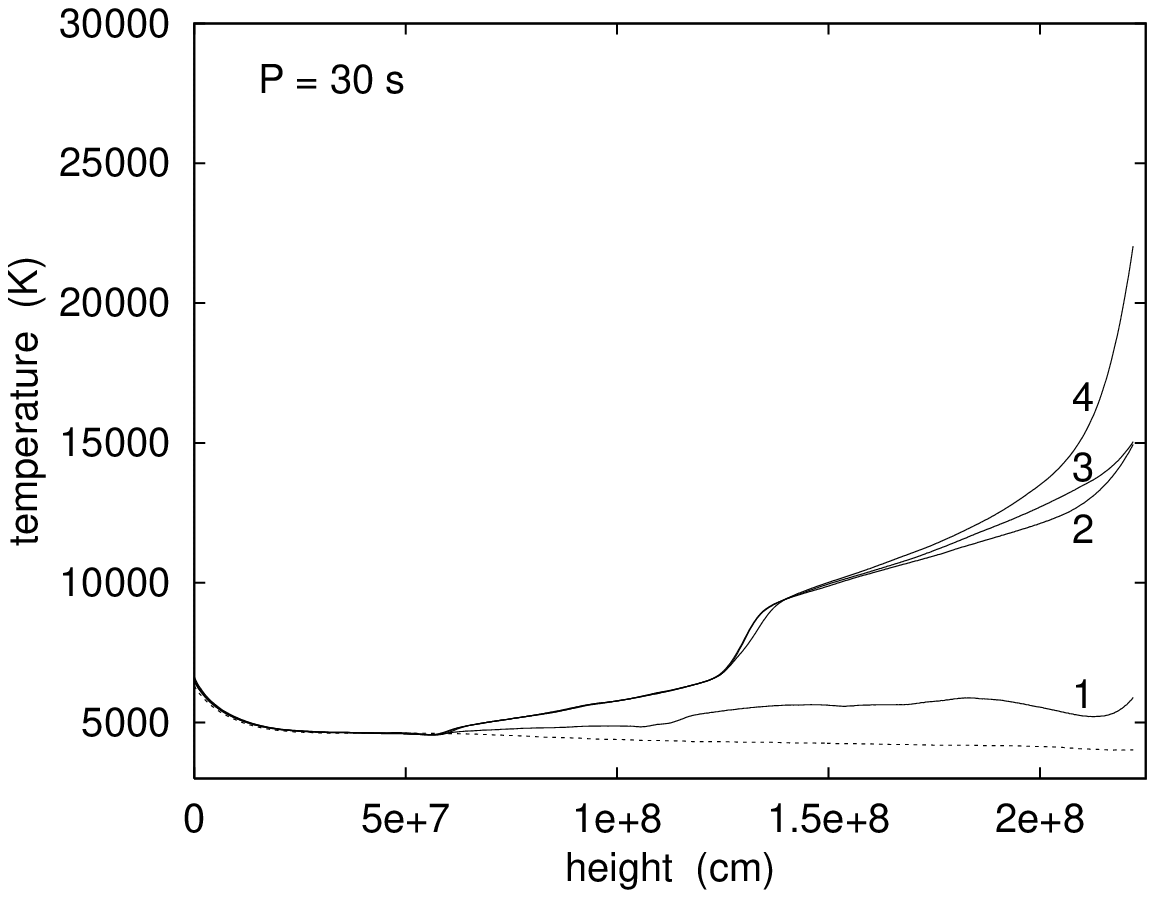,width=0.95 \textwidth,height=6.00cm}
\end{minipage}
\caption{
Mean temperatures as function of height at times indicated
$1 \, \hat{=} \, 500$~s, 
$2 \, \hat{=} \, 2000$~s,
$3 \, \hat{=} \, 3500$~s, 
$4 \, \hat{=} \,5000$~s for radiatively 
damped monochromatic waves with a wave energy flux of
$F_{\rm A} = 1\cdot10^8$~erg~cm$^{-2}$~s$^{-1}$ and a period of
$P=15$~s (left panel) as well as $P=30$~s (right panel). 
The temperature of the initial atmosphere is shown dotted.
} 
\end{figure*}
%%%%%%%%%%%%%%%%%%%%%%%%%%%%%%%%%%%%%%%%%%%%%%%%%%%%%%%%%%%%%%%%%%

Comparison of the two $P=15$~s spectra at $z=2000$~km shows that the 
adiabatic calculation in Fig.~1 has a much sharper 
resonance peak than that of Fig.~4. This is because in Fig.~1, 
the Fourier analysis is over the time span 500 to 2548~s, which 
still includes a considerable portion of the switch-on effect, 
while in Fig.~4 in the time between 1500 to 3548~s this effect is 
negligible. Another reason is that radiation damping, by 
decreasing the wave amplitudes, decreases the shock speeds and 
thus counteracts shock merging. Less shock merging always implies 
a larger amount of high-frequency power. 

\subsection{Effect of the generated chromosphere model, non-adiabatic 
case}
Including radiation changes the behaviour of the chromosphere 
model drastically. The temperature increase induced by shock 
heating is now counteracted by radiative cooling such that after some 
time, heating and cooling balance each other. Figure~5 shows the 
mean temperatures as function of height at 1500~s intervals both 
for the $P=15$ and 30~s waves. It is seen that in both cases 
after about 2000~s, a dynamically stable mean chromospheric 
temperature rise is established up to a height of $z=1400$~km. 
Here shock heating is balanced by H$^-$ and Mg~II~$k$ line 
cooling. At greater heights, where Mg~II cooling becomes 
inefficient and Ly$\alpha$ cooling becomes appreciable, a stable 
dynamical temperature distribution is achieved after about $t = 
5000$~s. 

It should be noted, however, that because of various reasons we 
do not consider this established temperature profile in the high 
chromosphere to be fully realistic. First we have used an arbitrarily 
selected monochromatic wave, second our treatment of radiation 
losses is rather incomplete both because of technical 
reasons and the small number of coolants considered. In 
addition, we have neglected the ionization energy of hydrogen in 
the energy budget of the wave and have omitted a fully 
consistent time-dependent treatment of the H ionization and 
recombination. In our treatment the ionization equilibrium 
follows the temperature variations instantaneously due to our 
solution of the time-dependent statistical rate equations for 
hydrogen.  As shown by Carlsson \& Stein (1992, 1994), the 
incorporation of the above mentioned effects significantly influences 
the temperature profiles particularly in the upper solar chromosphere. 

Let us now consider the acoustic wave spectra in Fig.~4 in view 
of the dynamic chromospheric models obtained. Figure~4 shows at 
height $z=2000$~km that the resonance oscillations in all three wave 
cases still persist after $t=3500$~s, whereas at the height 
$z=1300$~km, the spectra consist primarily of monochromatic 
components. Figure~6 gives the height of the shock merging events 
as function of time for the $P=15$ and 30~s waves. Comparison 
with Fig.~3 shows that, while the rate of shock merging for the 
30 s wave decreases with time similarly as in Fig.~3, 
that of the 15 s wave is very different. In the latter case the 
rate of shock merging events above $z=1400$~km does not appear 
to decrease with time. This shows that in the cases $P=15$ and 20~s,
a strong resonance contribution with very little or no 
contribution of the initial monochromatic excitation becomes 
established which persists with time.

As the shock merging events for the $P=30$~s wave occur 
increasingly close to the upper computational boundary at 
$z=2220$~km it can be expected that for this wave the spectrum
at $z=2000$~km in Fig.~4 will lose its resonance contributions
and eventually show only the monochromatic contribution. This 
supports the concept already discovered in Paper II of a 
critical period $P_{cr}\approx 25$~s below which the resonance 
oscillations are self-sustaining, while above that period the 
resonances die out. 

To summarize we find that for non-adiaba\-tic
(i.e., radiatively damped) monochromatic waves
the resonance behaviour is different compared to that of the 
adiabatic monochromatic waves. As the radiative cooling allows the atmosphere
to reach a dynamical steady state, shock merging processes 
take place now in an atmosphere of a fixed mean temperature.
Here waves with periods $P < P_{cr}$ find the sound speeds low enough
to catch up with one another in an atmospheric oscillation
kicked on by the initial switch-on effect. These shock merging events 
are able to perpetuate the resonance oscillations indefinitely. 
For periods $P>P_{cr}$, shock merging becomes increasingly
rare allowing the resonances to die out.

%%%%%%%%%%%%%%%%%%%%%%%%%%%%%%%%%%%%%%%%%%%%%%%%%%%%%%{6}%%%%%%%%%%% 
\begin{figure}[tb] 
\psfig{file=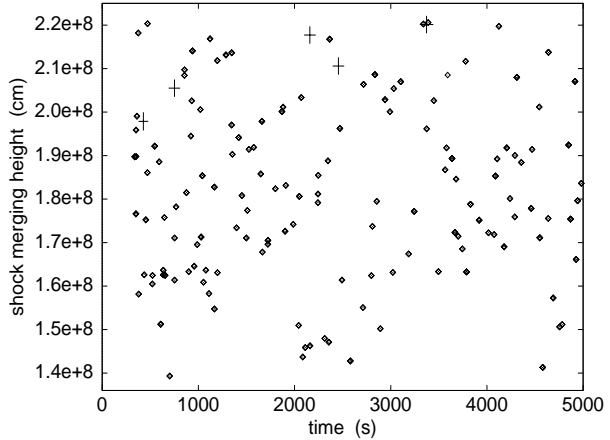,width=0.45 \textwidth,height=6.00cm} 
\caption{ Shock merging heights as function of time for 
radiatively damped monochromatic waves with an acoustic 
energy flux of $F_{\rm 
A} = 1\cdot10^8$~erg~cm$^{-2}$~s$^{-1}$ and a period of $P=15$~s 
(diamonds) as well as $P=30$~s (plus signs). } \end{figure} 
%%%%%%%%%%%%%%%%%%%%%%%%%%%%%%%%%%%%%%%%%%%%%%%%%%%%%%%%%%%%%%%%%% 
%
%%%%%%%%%%%%%%%%%%%%%%%%%%%%%%%%%%%%%%%%%%%%%%%%%%%%{7}%%%%%%%%%%%%%%%%%
\begin{figure*}[t]
\begin{minipage}[t]{0.45 \textwidth}
\psfig{file=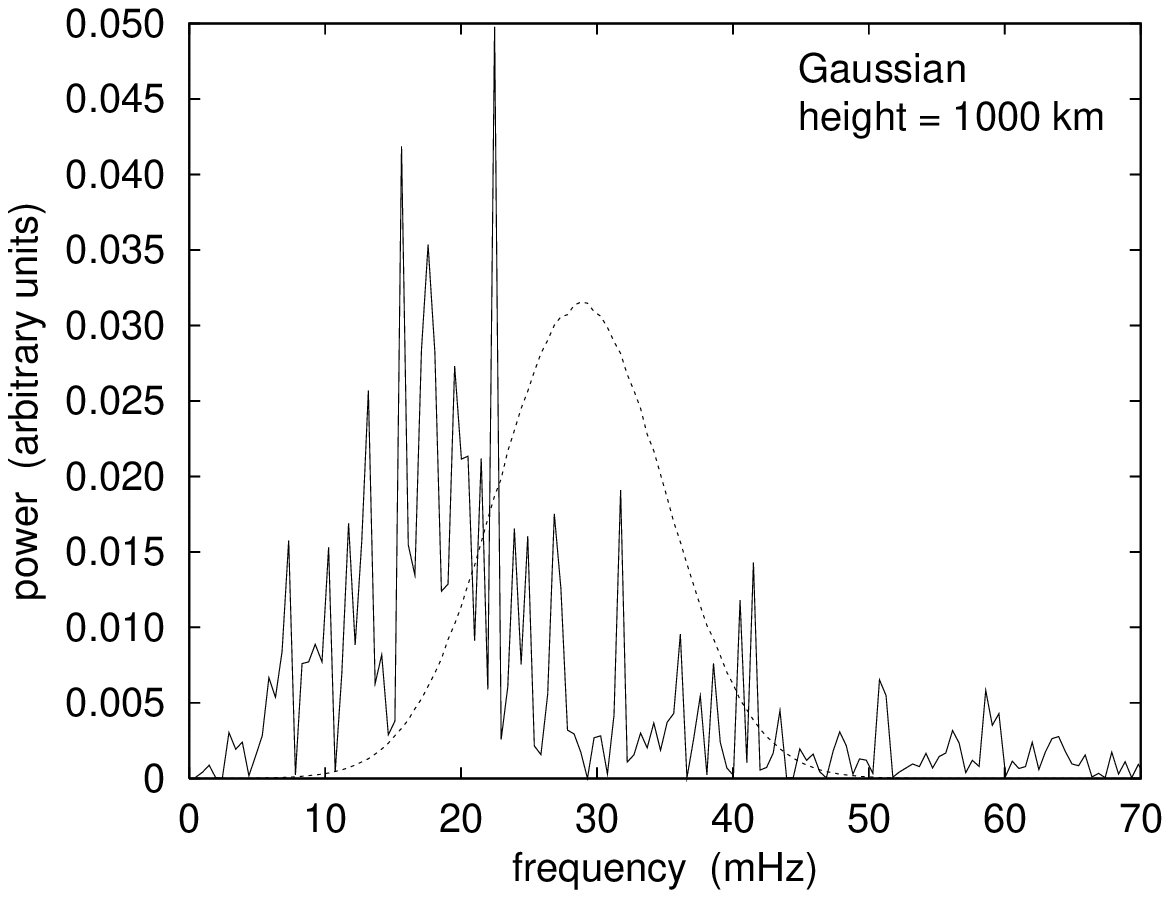,width=0.99 \textwidth,height=5.30cm}
\end{minipage}
\begin{minipage}[t]{0.45 \textwidth}
\psfig{file=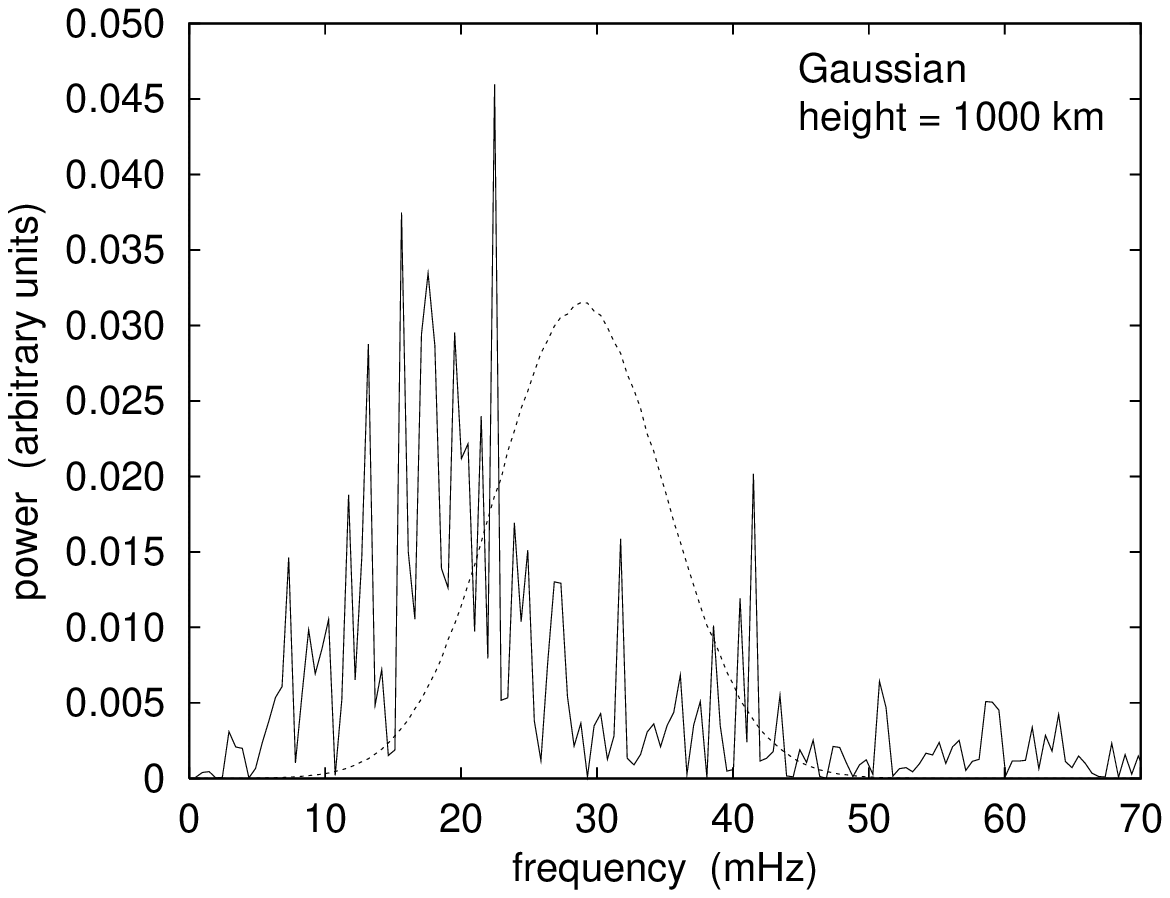,width=0.99 \textwidth,height=5.30cm}
\end{minipage}
%%%%%%%%%%%%%%%%%%%%%%%%%%%%%%%%%%%%%%%
\begin{minipage}[t]{0.45 \textwidth}
\psfig{file=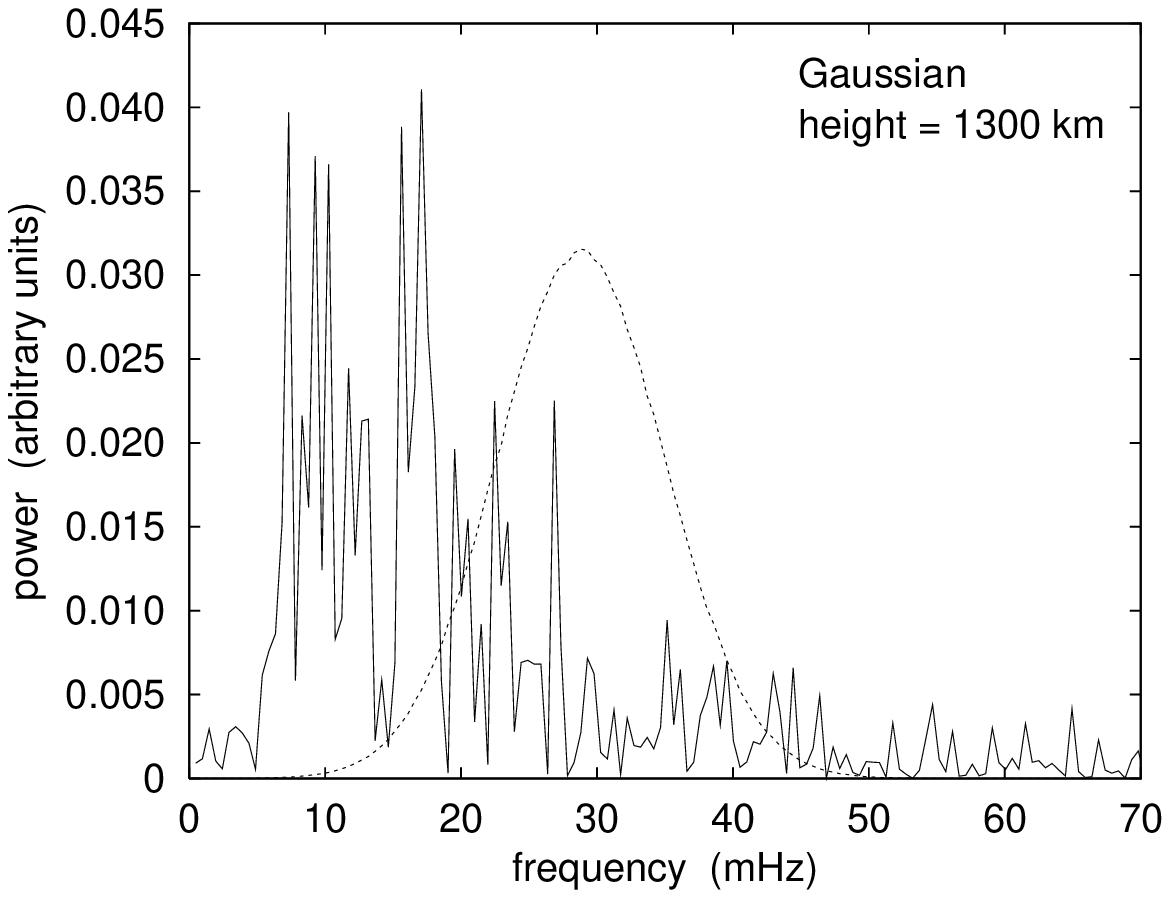,width=0.99 \textwidth,height=5.30cm}
\end{minipage}
\begin{minipage}[t]{0.45 \textwidth}
\psfig{file=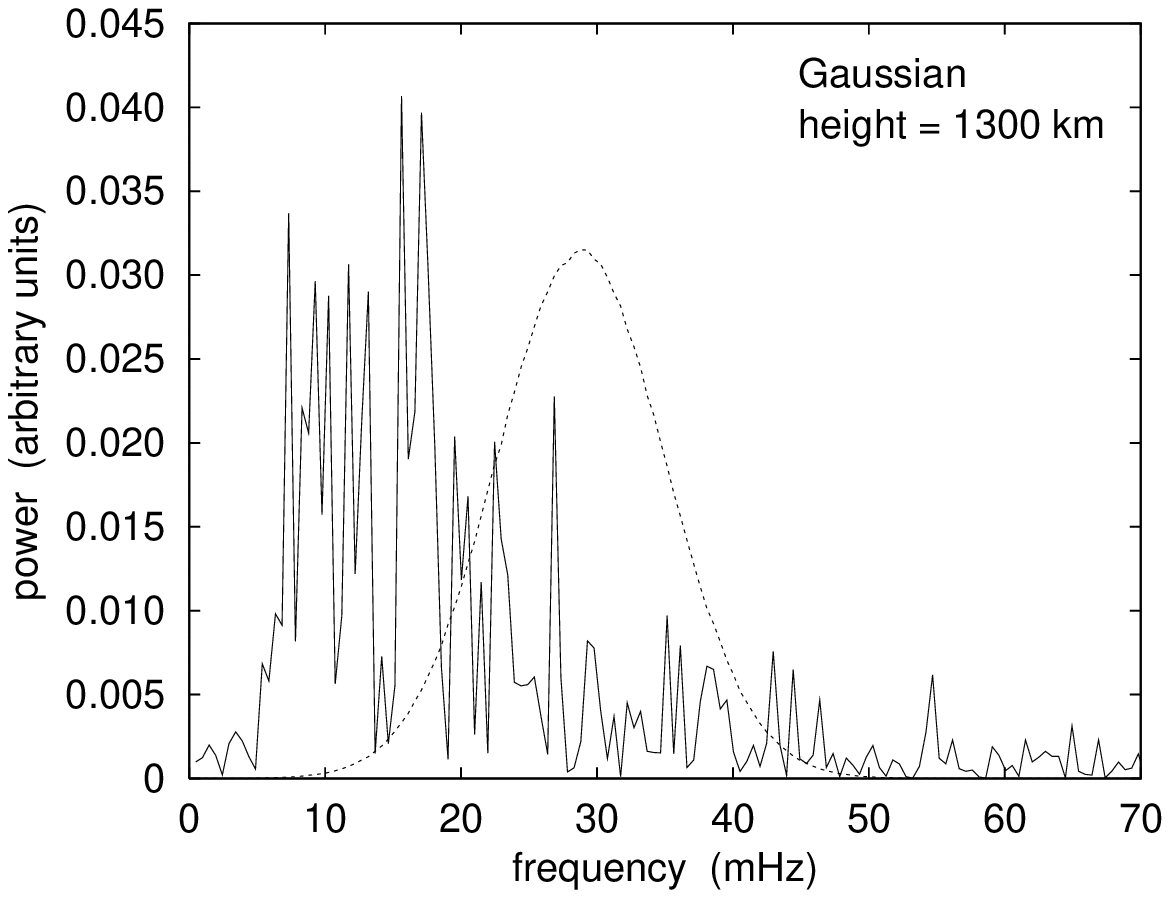,width=0.99 \textwidth,height=5.30cm}
\end{minipage}
%%%%%%%%%%%%%%%%%%%%%%%%%%%%%%%%%%%%%%%
\begin{minipage}[t]{0.45 \textwidth}
\psfig{file=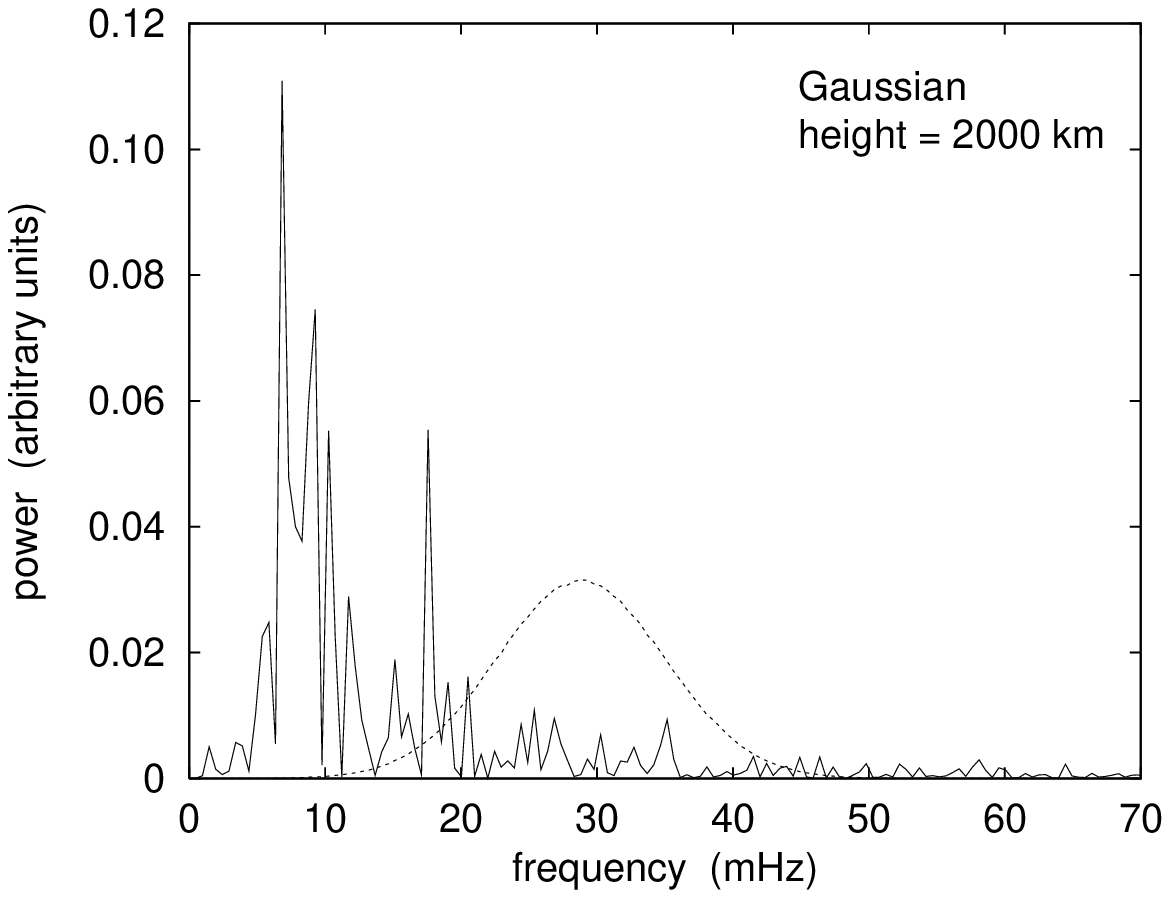,width=0.99 \textwidth,height=5.30cm}
\end{minipage}
\begin{minipage}[t]{0.45 \textwidth}
\psfig{file=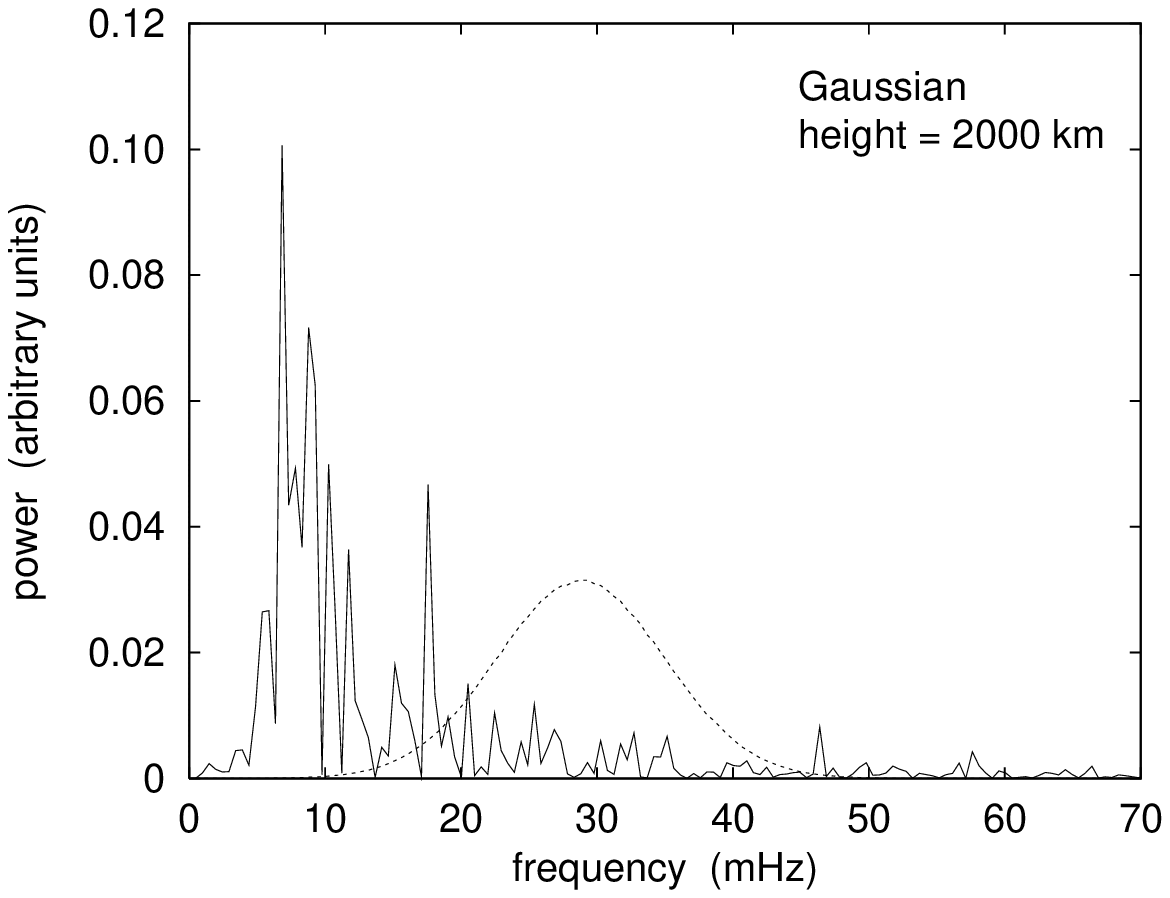,width=0.99 \textwidth,height=5.30cm}
\end{minipage}
\caption{
Power spectra at 0 (dotted), 1000, 1300, 2000 km height (from 
top to bottom), Fourier analysing the time span 500 - 2548 s (left 
column) and 2500 - 4548 s (right column) 
for an atmosphere excited at height $z=0$~km by a 
Gaussian acoustic frequency spectrum with a flux of
$F_{\rm A} = 1\cdot10^8$~erg~cm$^{-2}$~s$^{-1}$ and a central maximum 
at period $P=35$~s. The wave calculation is 
non-adiabatic.
}
\end{figure*}
%%%%%%%%%%%%%%%%%%%%%%%%%%%%%%%%%%%%%%%%%%%%%%%%%%%%%%%%%%%%%%%%%%%%%%

\subsection{Excitation by acoustic spectra for non-adiabatic waves}

We now consider the excitation of the solar atmosphere by 
various acoustic wave spectra with radiation damping included. 
For this exploratory investigation we select only two of the 
four excitation spectra considered in Paper II. The first one is a 
Gaussian spectrum with a central peak at frequency $\omega_C = 
2\pi / 35$~Hz, extending from 10 to 50 mHz, the second is a stochastic 
spectrum extending from 5 to 55 mHz (see Paper~II). As in Paper II 
we represent the input spectrum by 101 partial waves
equidistantly spaced in frequency with $\Delta\nu = 0.5$~mHz. The 
lowest frequency point is about a factor of 0.7 below the cut-off 
frequency $\nu_A \approx 5$~mHz, the highest about a factor of 
11 larger. 
%%%%%%%%%%%%%%%%%%%%%%%%%%%%%%%%%%%%%%%%%%%%%%%%%%%{8}%%%%%%%%%%%%%%%%%%
\begin{figure*}[t]
\begin{minipage}[t]{0.45 \textwidth}
\psfig{file=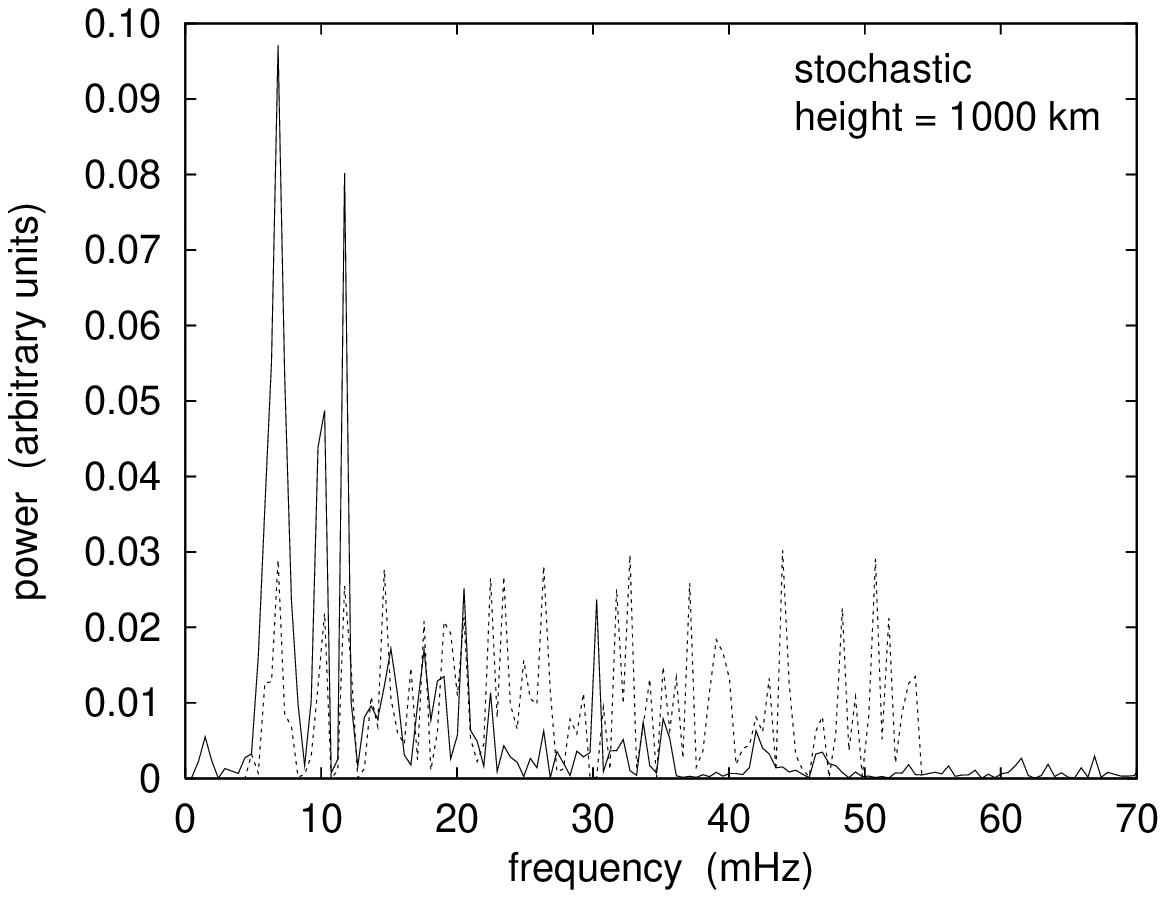,width=0.99 \textwidth,height=5.30cm}
\end{minipage}
\begin{minipage}[t]{0.45 \textwidth}
\psfig{file=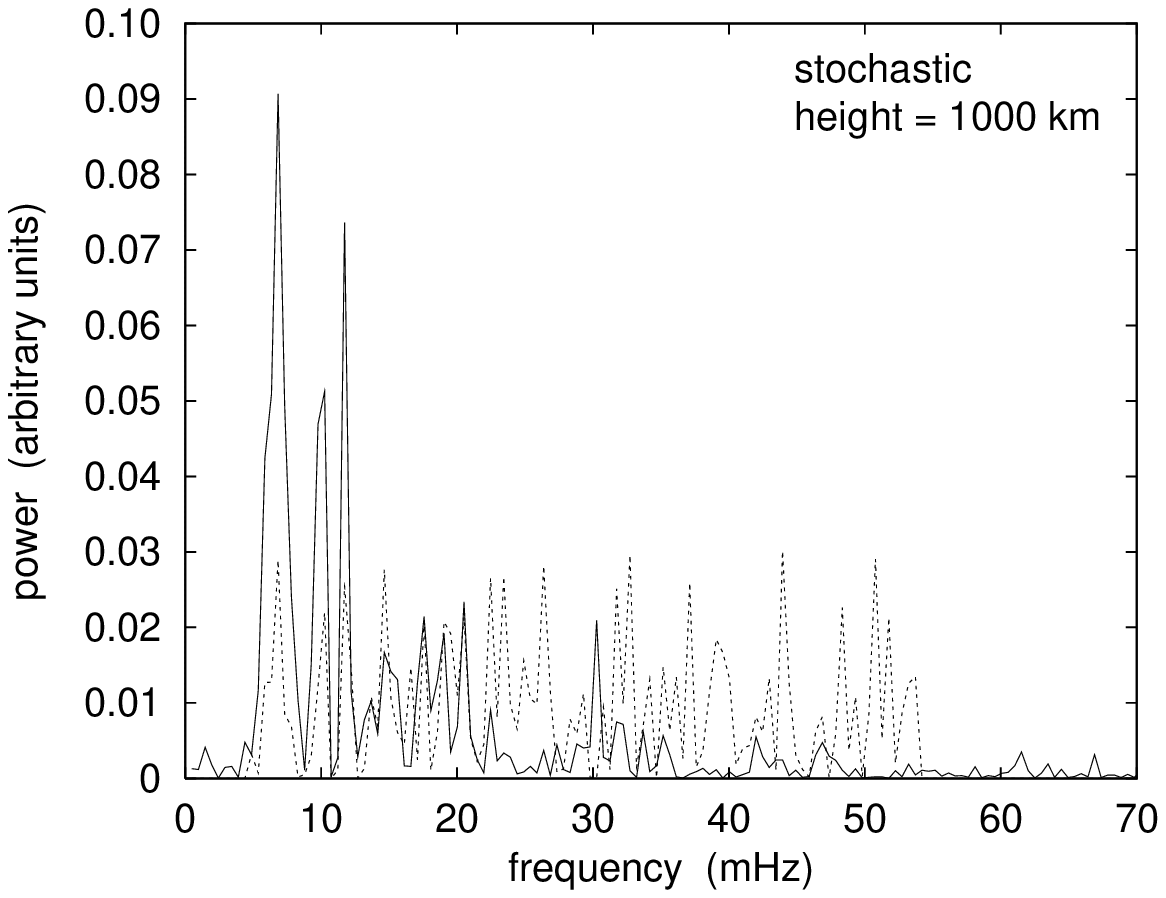,width=0.99 \textwidth,height=5.30cm}
\end{minipage}
%%%%%%%%%%%%%%%%%%%%%%%%%%%%%%%%%%%%%%%
\begin{minipage}[t]{0.45 \textwidth}
\psfig{file=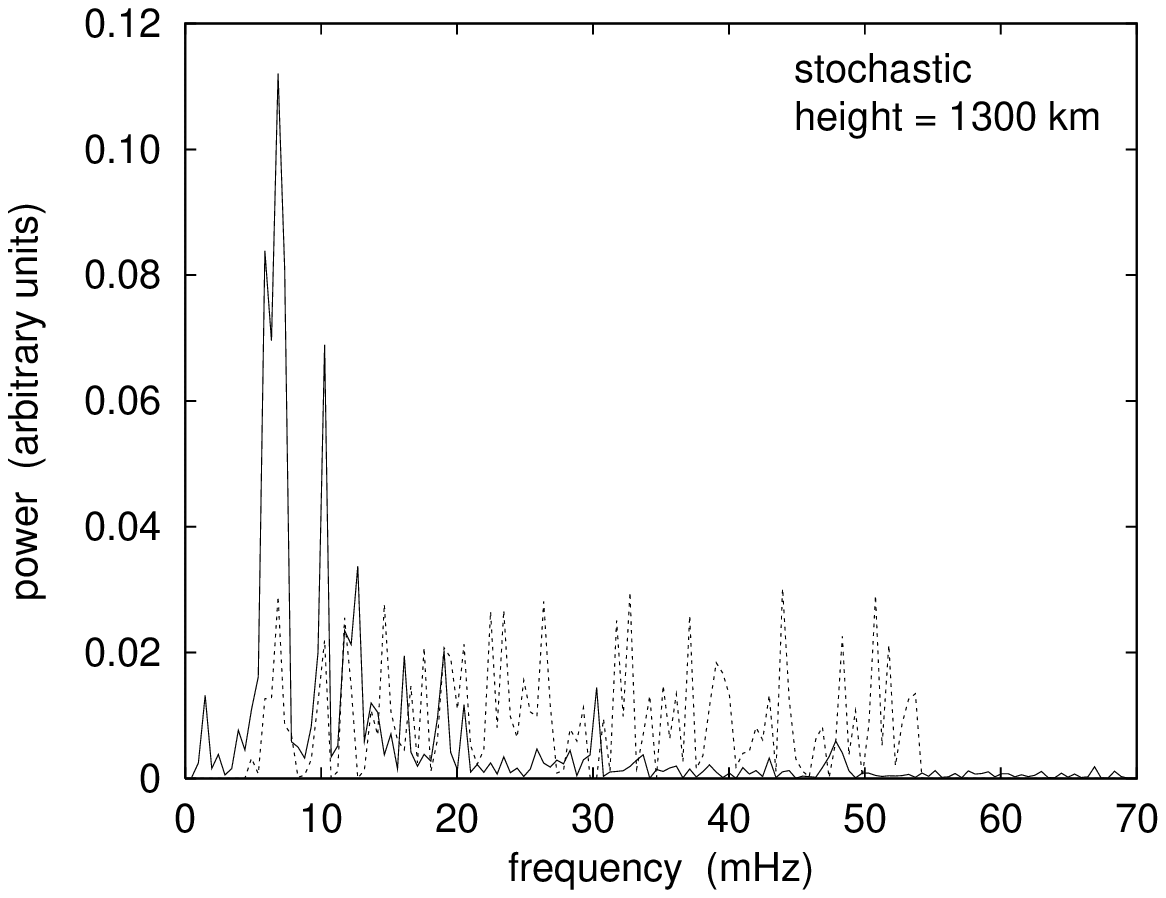,width=0.99 \textwidth,height=5.30cm}
\end{minipage}
\begin{minipage}[t]{0.45 \textwidth}
\psfig{file=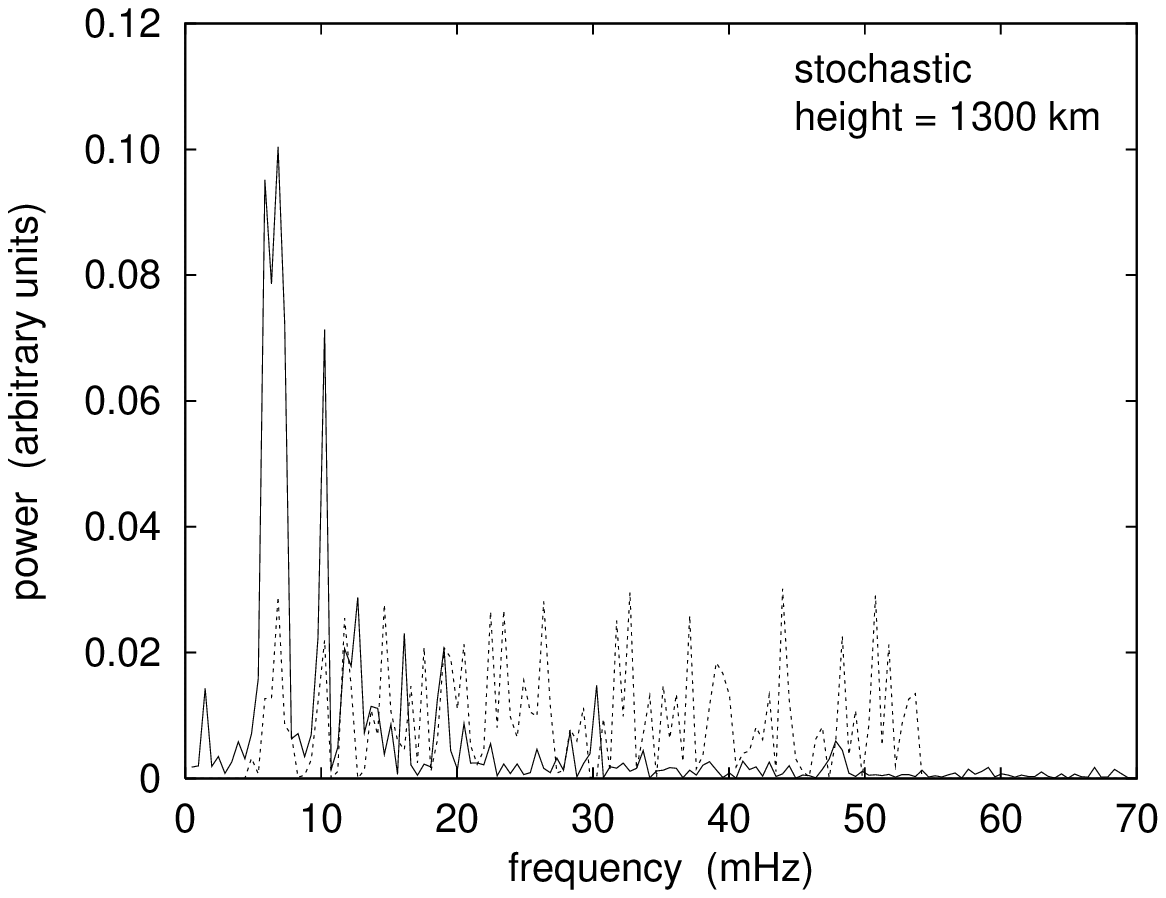,width=0.99 \textwidth,height=5.30cm}
\end{minipage}
%%%%%%%%%%%%%%%%%%%%%%%%%%%%%%%%%%%%%%%
\begin{minipage}[t]{0.45 \textwidth}
\psfig{file=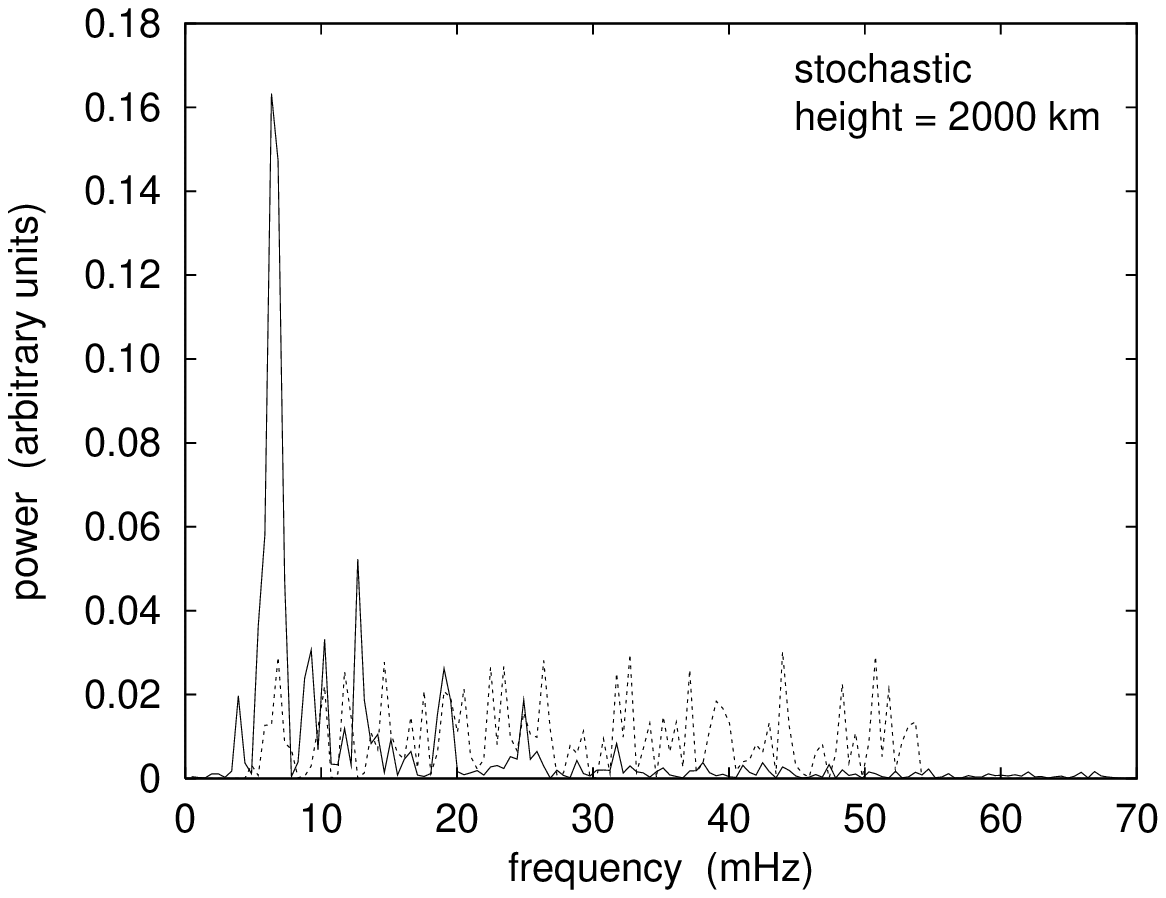,width=0.99 \textwidth,height=5.30cm}
\end{minipage}
\begin{minipage}[t]{0.45 \textwidth}
\psfig{file=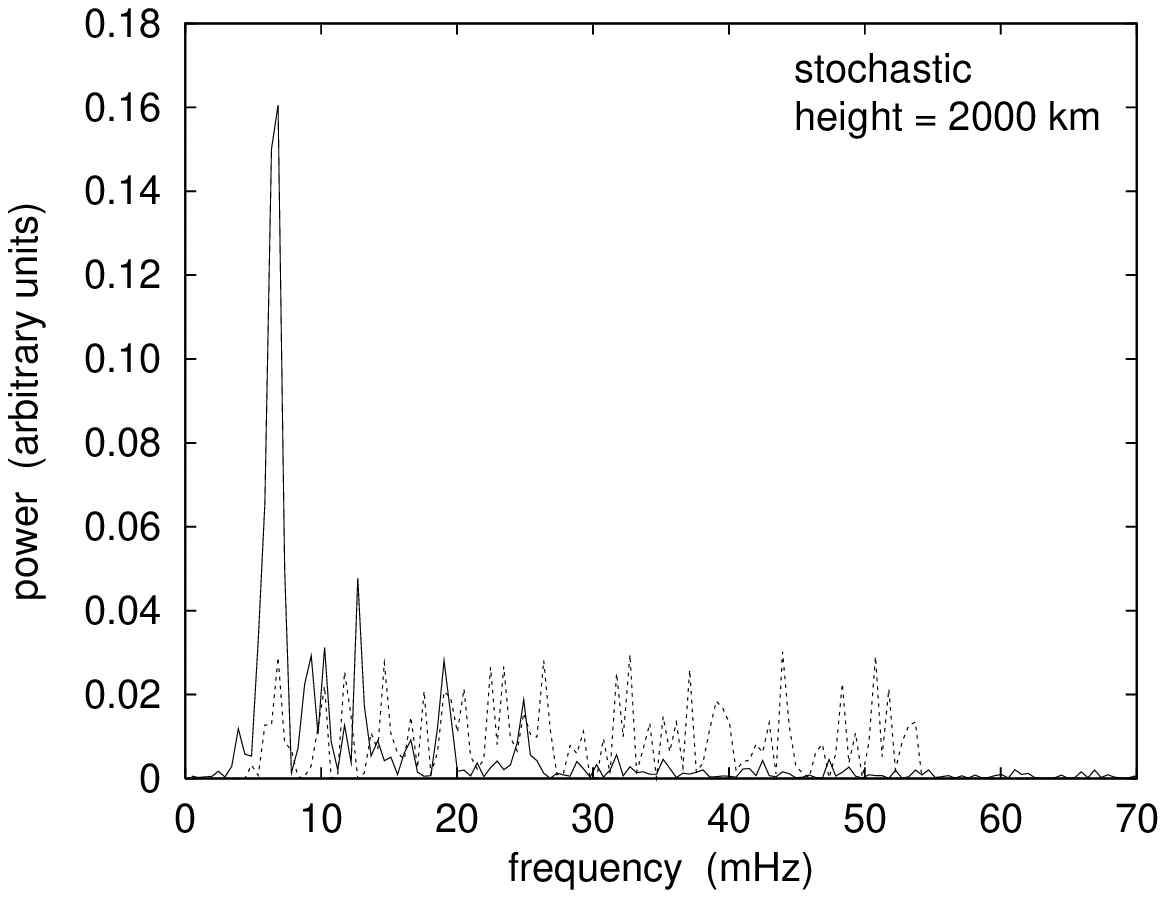,width=0.99 \textwidth,height=5.30cm}
\end{minipage}
\caption{
Same as Fig.~7, however for a stochastic acoustic frequency spectrum.
}
\end{figure*}
%%%%%%%%%%%%%%%%%%%%%%%%%%%%%%%%%%%%%%%%%%%%%%%%%%%%%%%%%%%%%%%%%%%%%%

For a comparison with our monochromatic studies we take the same 
total energy flux
$F_{\rm A} = 1\cdot10^8$~erg~cm$^{-2}$~s$^{-1}$ for the
spectra. The initial acoustic wave spectra which excite 
the solar atmosphere model at height $z=0$~km are shown (dotted) 
in all panels of Figs.~7 and 8. These figures show (drawn) the 
acoustic wave spectra at heights $z=1000$, $1300$ and $2000$~km 
(from top to bottom). Furthermore, different timespans for the Fourier 
analysis are taken. We selected the time intervals 500 - 2548 s (left 
column) and 2500 - 4548 s (right 
column). There are two striking results from these spectra. 
First, it is seen that in both wave cases there is a strong and 
with height progressing shift of the spectral power from the 
initially specified one to one which shows a predominant
contribution in the 6-8 mHz (3 min) range. Comparing Figs.~7 and 
8 it is seen that despite the very different initial shapes 
of the two spectral cases, they become increasingly similar for 
greater height. At 2000 km, one has almost pure 3 min 
spectra in both cases.
 
The second striking feature of Figs.~7 and 8 is that the spectra 
at a given height apparently do not vary with time. Comparing 
the spectra computed from the time interval 500 - 2548~s with those of 
2500 - 4548~s at a given height shows that there is 
essentially no change. This is very different from the 
monochromatic adiabatic results where the resonance oscillations 
are transient and die out with time. 

%%%%%%%%%%%%%%%%%%%%%%%%%%%%%%%%%%%%%%%%%%%%%%%%%%%%%{9}%%%%%%%%%%%%%%%%
\begin{figure*}[t]
\begin{minipage}[t]{0.45 \textwidth}
\psfig{file=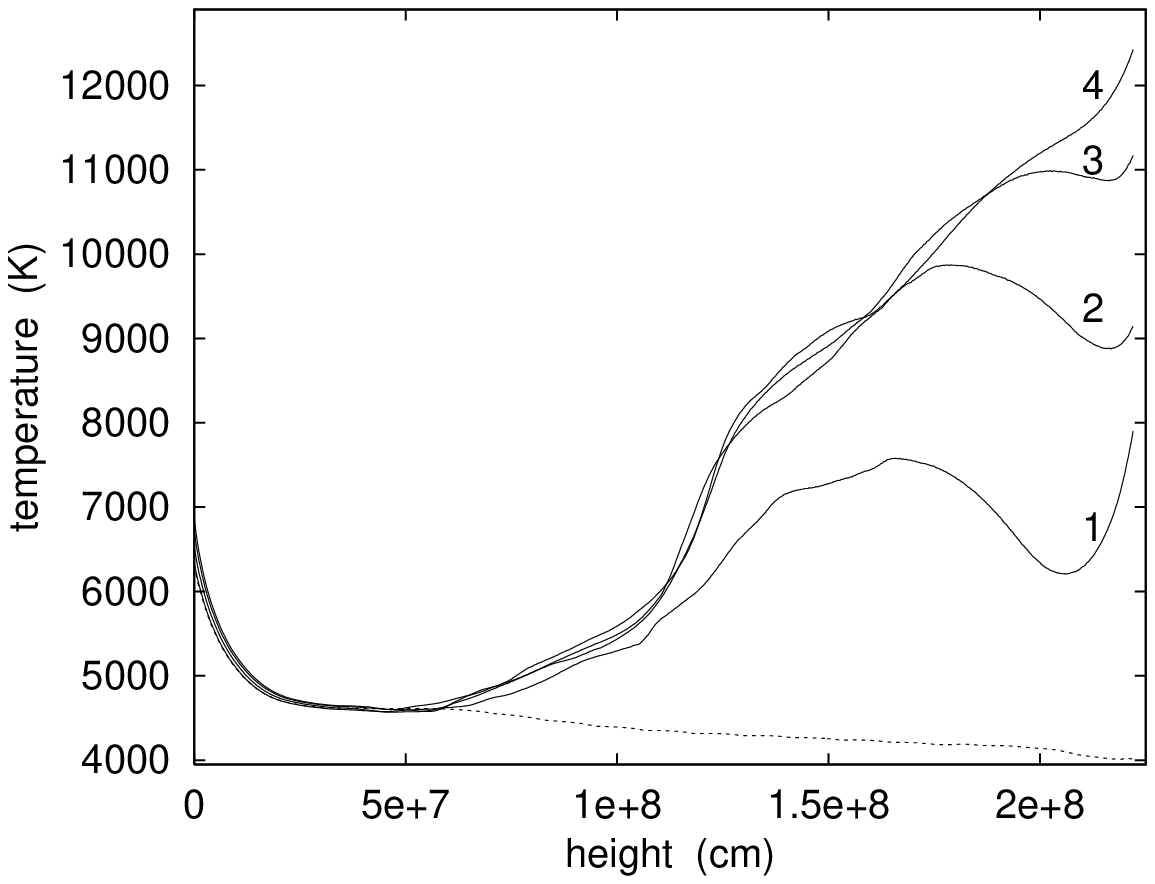,width=0.95 \textwidth,height=5.90cm}
\end{minipage}
\begin{minipage}[t]{0.45 \textwidth}
\psfig{file=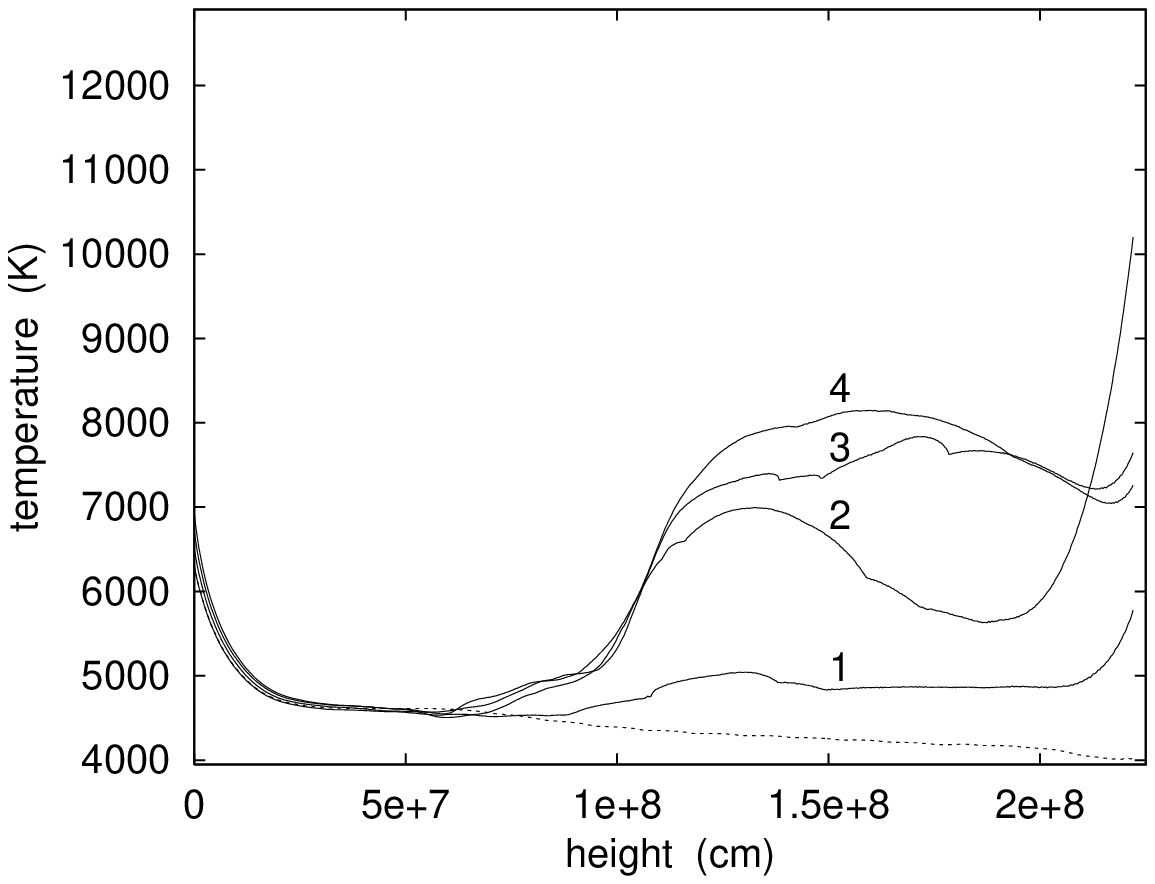,width=0.95 \textwidth,height=5.90cm}
\end{minipage}
\caption{
Mean temperatures as function of height at times indicated
$1 \, \hat{=} \, 500$~s, $2 \, \hat{=} \, 2000$~s, 
$3 \, \hat{=} \, 3500$~s, $4 \, \hat{=} \, 5000$~s for 
non-adiabatic acoustic wave spectra, i.e.~the Gaussian (left panel) and
the stochastic one (right panel).
The temperature of the initial atmosphere is shown dotted.
} 
\end{figure*}
%%%%%%%%%%%%%%%%%%%%%%%%%%%%%%%%%%%%%%%%%%%{10}%%%%%%%%%%%%%%%%%%%%%%%%%%
\begin{figure*}[t]
\begin{minipage}[t]{0.45 \textwidth}
\psfig{file=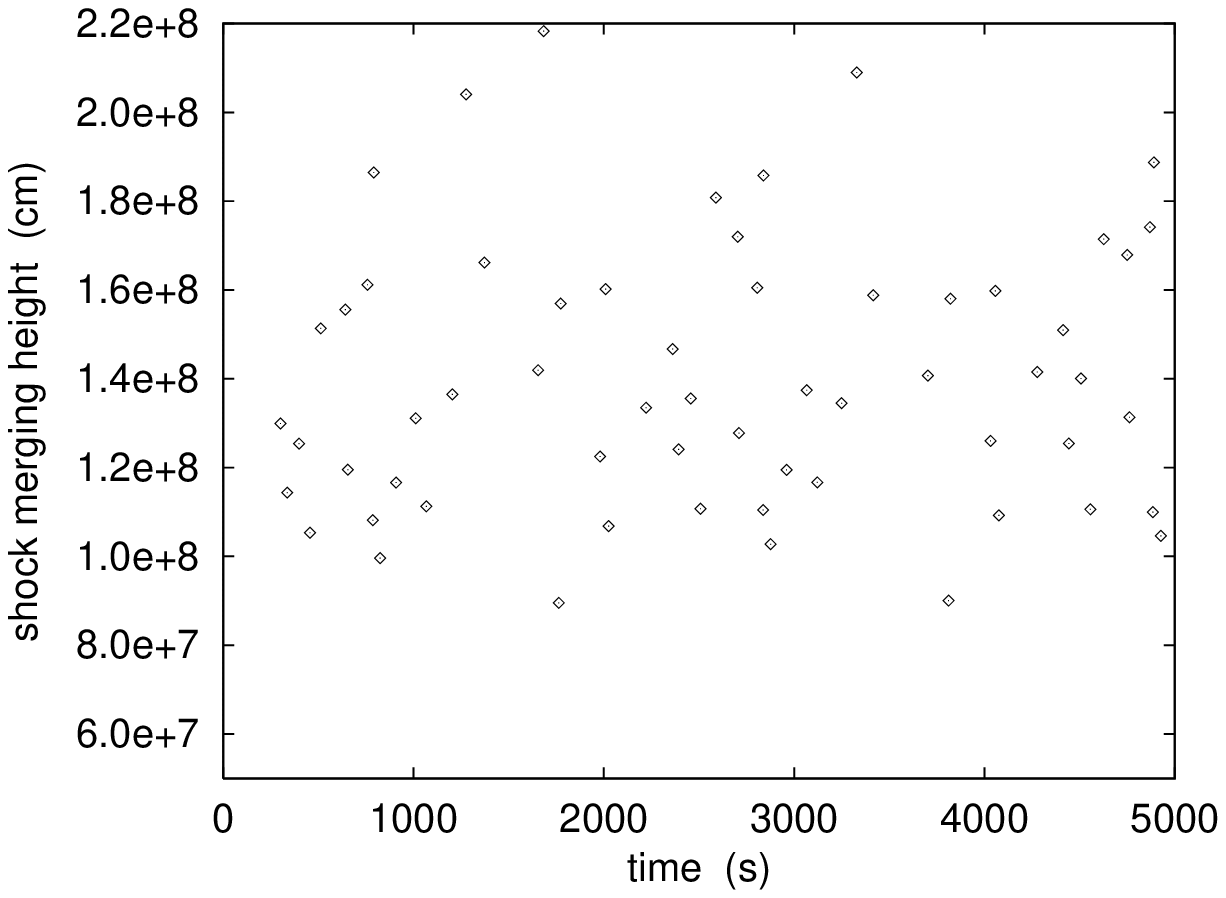,width=0.95 \textwidth,height=5.90cm}
\end{minipage}
\begin{minipage}[t]{0.45 \textwidth}
\psfig{file=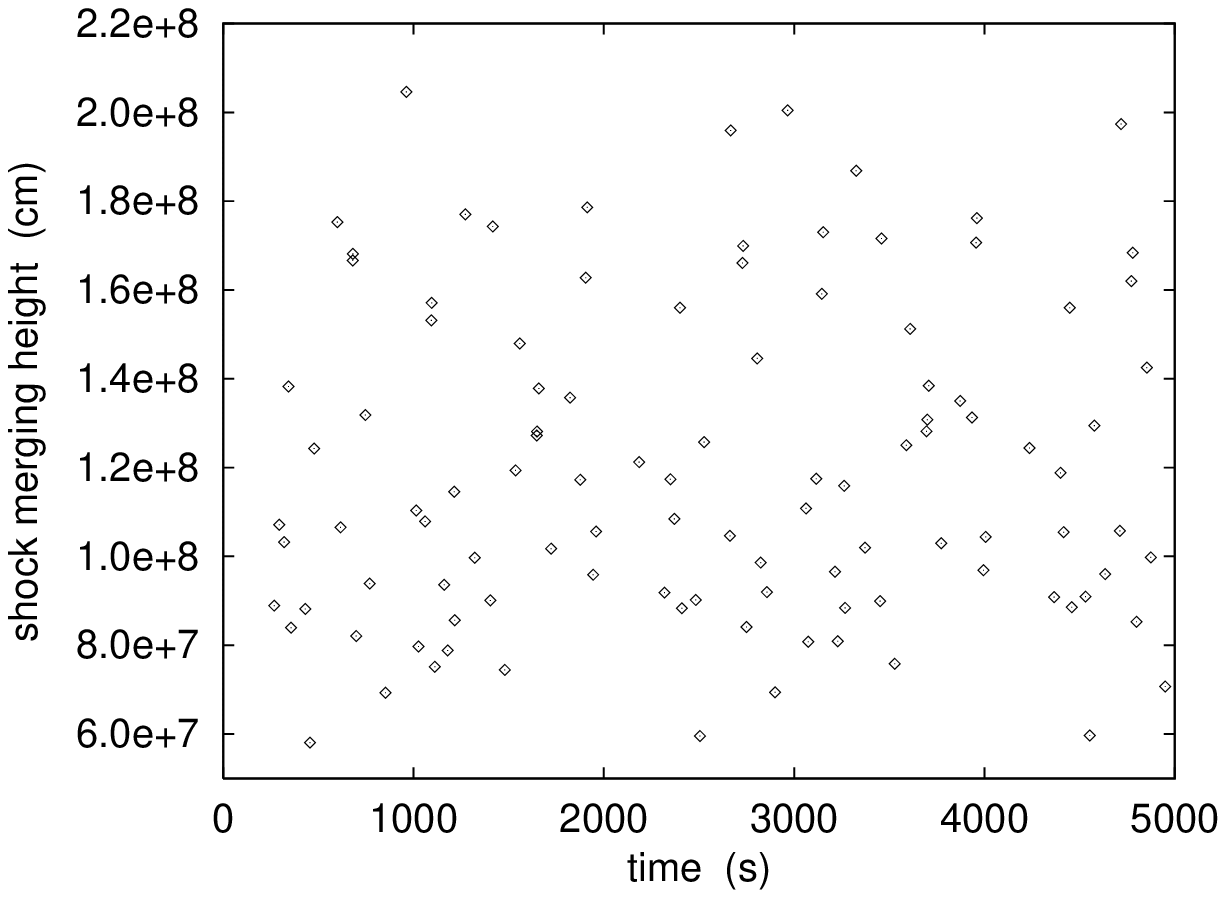,width=0.95 \textwidth,height=5.90cm}
\end{minipage}
\caption{
Shock merging heights as function of time
for the Gaussian acoustic wave spectrum (left panel) and 
the stochastic spectrum (right panel).               
} 
\end{figure*}
%%%%%%%%%%%%%%%%%%%%%%%%%%%%%%%%%%%%%%%%%%%%%%%%%%%%%%%%%%%%%%%%%%%%%%
As found in our monochromatic studies above, a combination of two 
conditions will lead to a continuous presence of
resonance oscillations: the atmosphere must reach a dynamical 
equilibrium and the shock merging events must go on 
indefinitely. Both conditions are met in an excitation 
with frequency spectra. Figure~9 shows the mean temperatures of the  
atmosphere as function of time for the two wave cases considered.
It is seen that like in Fig.~5, a dynamical steady state temperature 
distribution is established below 1400 km height after 2000 s and 
somewhat later after 3500 s at greater height. For the second 
condition, Fig.~10 shows the plots of the shock merging heights 
of the two wave cases considered.
It is seen that for the Gaussian spectrum,
shock merging goes on indefinitely with time in the height interval
between 1000 km and the computational boundary at 2220 km. For the 
stochastic wave case, incessant shock merging occurs
in the range 600 to 2000~km. 

Because shock merging implies the presence of shocks, and 
because the initial acoustic frequency spectra have relatively 
small amplitude with no shocks present, it is clear that the 
shocks have to be generated during the wave propagation. Figure~11 
shows for the stochastic case that different to the situation of 
monochromatic waves, where all shocks form close to 400 km 
height (see Fig.~2), the shocks in the case of an acoustic 
spectrum are formed in a wide range between 300 and 1100 km 
height. Above 1100 km height, the waves are all 
sawtooth-type shock waves. 

\subsection{Shock merging and the generation of 3 min oscillations}
At this point we want to discuss in more detail the relation 
between shock merging and the generation of 3 min oscillations.
In the work of Lites et al.~(1993) as well as 
Cheng \& Yi (1996) it is argued on basis of observations
that in the solar atmosphere the 3 min signal is already present 
in the lower photosphere and that therefore shock merging 
is not an important process for the generation of 3 min 
oscillations. We believe that contrary to this view, our present 
calculations clearly show that shock merging is a powerful method
for generating 3 min oscillations.
%%%%%%%%%%%%%%%%%%%%%%%%%%%%%%%%%%%%%%%%%%%%%%%%%%%%%{11}%%%%%%%%%%%
\begin{figure}[t]
\psfig{file=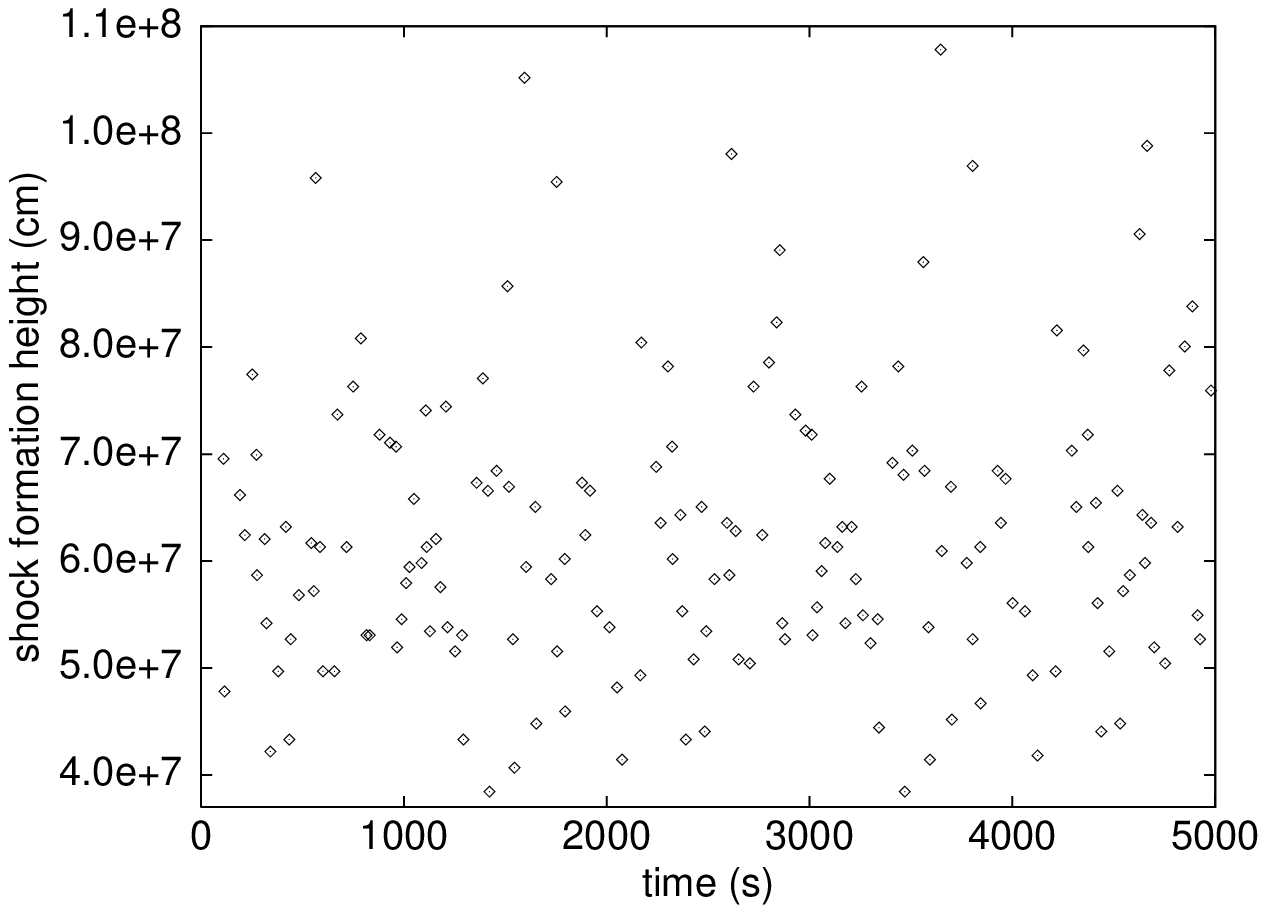,width=0.45 \textwidth,height=5.90cm}
\caption{
Shock formation heights as function of time
for a radiatively damped stochastic wave spectrum of
an acoustic energy flux of 
$F_{\rm A} = 1\cdot10^8$~erg~cm$^{-2}$~s$^{-1}$.
}
\end{figure}
%%%%%%%%%%%%%%%%%%%%%%%%%%%%%%%%%%%%%%%%%%%%%%%%%%%%%%%%%%%%%%%%%

Our monochromatic calculations indicate that if the
wave periods are not too small, the resonance oscillations die
out. However, our calculations with acoustic wave spectra
show a fundamentally different
behaviour. One might think that in an acoustic spectrum
(cf.~Fig.~2) the individual partial waves simply form a
wavetrain of sawtooth waves and thus do not generate low
frequency oscillations. However, the stochastic nature of the
superposition of partial waves in an acoustic spectrum
perpetually leads to large amplitude fluctuations which
invariably generate shocks of different wave period and
strength. In addition, note that stochastic superpositions
lead to perpetual resonances even in the small amplitude case
(see Fig.~14 of Paper I).

The propagation of the shocks in an oscillating atmosphere
subsequently results in shock merging events which in turn
drive the atmospheric oscillations. Thus acoustic wave
spectra will invariably lead to perpetual resonance
oscillations. As in our assumed input spectra the
3 min signal is absent, these resonance oscillations must have been
generated by the atmosphere in response to the wave excitation.
As the acoustic spectrum
calculations of Musielak et al.~(1994) do not show a
prominent 3 min contribution at the top of the convection zone,
we consequently conclude that the 3 min band present in
the Lites et~al.\ observations must have been added later by
hydrodynamic effects in the solar atmosphere. 
This conjecture will be investigated
further in the subsequent paper V.

\subsection{Acoustic spectra and the dynamic structure of the 
atmosphere} A very important result found by comparing the two 
cases of Fig.~9 is that the mean temperature of the atmosphere 
in a dynamical steady state strongly depends on the initial wave 
spectrum. This result also applies to the temperature behaviour of 
Fig.~5 for the unrealistic case of monochromatic waves. We 
therefore conclude that the structure of stellar chromospheres is 
intimately determined by the initial acoustic wave spectrum, 
generated in the convection zone. It thus should be 
possible to infer the initial wave spectrum by detailed 
simulations by comparing theoretical spectra (see e.g.~Figs.~7 
and 8) and observed spectra at various atmospheric heights.
 
\section{Conclusions}
From our adiabatic and non-adiabatic wave calculations for a 
solar atmosphere model, excited at the bottom by large 
amplitude monochromatic 
waves as well as by acoustic frequency spectra, employing a 
realistic mechanical flux of
$F_{\rm A} = 1\cdot10^8$~erg~cm$^{-2}$~s$^{-1}$,
we draw the following conclusions: 

\smallskip
\noindent 1. Adiabatic wave calculations, because of the 
unbalanced heating, invariably lead to chromospheric temperature 
plateaus with perpetually rising mean temperatures. This 
time-depend\-ent growth of the mean temperature makes it 
increasingly difficult for shock merging to occur. The 
consequence is that the 3 min type resonances, kicked on by the 
shock merging events, die out and after some time only the 
monochromatic signal survives.  

\smallskip
\noindent 2. Non-adiabatic calculations, where the radiative 
cooling by NLTE H$^-$ continua as well as by Mg II $k$ and H$^-$
Ly$\alpha$ lines are considered, invariably establish a 
dynamically generated mean chromospheric temperature 
distribution, which at heights below 1400 km is established 
after 2000 s and at greater height after about 3500 s. 

\smallskip
\noindent 3. For excitation by non-adiabatic monochromatic 
waves, as already noted in Paper II, a critical frequency 
$\nu_{cr}\sim 1/25$~Hz is found, which separates domains of 
drastically different resonance behaviour. For frequencies 
$\nu<\nu_{cr}$, the atmospheric resonance decays, leaving behind 
only the monochromatic signal, while for $\nu>\nu_{cr}$ 
the resonance oscillation persists and is kicked on by shock 
merging. For wave periods $P < 25$~s, the shock merging occurs 
undiminished in the height range 1400 to 2200 km. Here the 
resonances appear self-sustaining. 

\smallskip
\noindent 4. Using two different acoustic wave spectra
(Gaussian and stochastic) to excite the atmosphere at height 
$z=0$~km, we found that in non-adiabatic wave calculations, 
the spectra showed a characteristic shift towards lower 
frequency such that at height $z=2000$~km essentially only a 
pure 3 min band exists. 

\smallskip
\noindent 5.  We also found that 
at any given height the acoustic spectra, after 
an initial time of 500 s, did no longer change with time.

\smallskip
\noindent 6. In the height interval 1000 - 2200 km for the 
Gaussian and 600 - 2000 km for the stochastic spectrum, shock 
merging occurs persistently with time. This self-sustaining
shock-merging behaviour is responsible for the fact that there 
is a 3 min resonance band in the acoustic spectra, which becomes
increasingly pronounced with atmospheric height.

\smallskip
\noindent 7. From our calculations we surmise that the
3 min component observed by Lites et al.~(1993)
in a low lying iron line, which is not found in the acoustic 
spectrum generated in the convection zone (Musielak et al.~1994),
has been added later by the solar atmosphere as response
to the propagation of the acoustic wave spectrum.

\smallskip
\noindent 8. We find that the dynamically generated mean 
chromospheric temperature structure is strongly dependent on the 
assumed initial acoustic wave spectrum. This indicates that 
detailed wave simulations and height--dependent spectral 
observations will allow to empirically determine the velocity 
fluctuations in the acoustic wave generation region. This could 
be an important independent check of existing convection and
sound generation simulations.

\smallskip
\noindent 9. The appreciable atmospheric expansions found in 
Paper II did not occur in the present simulations because we now 
use an Eulerian code. 

\bigskip
%------------------------------- Acknowledgements --------------------------
\noindent
\begin{acknowledgements}
We thank the Deutsche Forschungsgemeinschaft for grant UL 57/22-1
and NATO for grant CRG-910058. This work was also partially 
supported by the NASA Astrophysics Theory Program (grant number 
NAG5-3027) to the University of Alabama in Huntsville (P.U., M.C.).
We also gratefully acknowledge valuable comments by 
F.-L.~Deubner and S.~Steffens 
on an earlier version of the manuscript.
\end{acknowledgements}
%-------------------------------- References -------------------------------


\begin{thebibliography}{qqq}
 
\bibitem{aab} Buchholz B., Hauschildt P.H., Rammacher W., Ulmschneider 
                   P., 1994, A\&A 285, 987 

\bibitem{aac} Carlsson M., Stein R.F., 1992, ApJ 397, L59 

\bibitem{aad} Carlsson M., Stein R.F., 1994. 
In: Carlsson M.~(ed.) Proc.~Oslo Mini-Workshop at 
Inst.~of Theoretical Astrophysics,
Chromospheric Dynamics, p.~47

\bibitem{aae} Cheng Q.-Q., Yi Z., 1996, A\&A 313, 971

\bibitem{aaf} Cuntz M., 1987, A\&A 188, L5 

\bibitem{aag} Cuntz M., Ulmschneider P., 1988, A\&A 193, 119

\bibitem{xxx} Deubner F.-L., Reichling M., Langhanki R., 1988. 
  In: Christensen-Dalsgaard J., Frandsen S.~(eds.) IAU Symp.~123,
       Advances in Helio- and Asteroseis\-mology, p.~439

\bibitem{aah} Deubner F.-L., 1991. 
  In: Ulmschneider P., Priest E.R., Rosner R.~(eds.) 
    Mechanisms of Chromospheric and Coronal 
    Heating. Springer, Berlin, p.~6
 
\bibitem{aai} Fleck B., Deubner F.-L., 1989, A\&A 224, 245

\bibitem{aaj} Fleck B., Schmitz F., 1991, A\&A 250, 235
 
\bibitem{aak} Fleck B., Schmitz F., 1993, A\&A 273, 671

\bibitem{aal} Judge P.G., 1990, ApJ 348, 279
 
\bibitem{aam} Kalkofen W., Rossi P., Bodo G., Massaglia S., 1994, A\&A 284, 976
 
\bibitem{aan} Leibacher J.W., Stein R.F., 1981. 
    In: Jordan S.~(ed.) The Sun as a Star. NASA SP-450, p.~263

\bibitem{aao} Lites B.W., Rutten R.J., Kalkofen W., 1993, ApJ 414, 345
 
\bibitem{aap} Musielak Z.E., Rosner R., Stein R.F., Ulmschneider P., 
    1994, ApJ 423, 474

\bibitem{aaq} Rammacher W.,  Ulmschneider P., 1992, A\&A 253, 586

\bibitem{aar} Rossi P., Kalkofen W., Bodo G., Massaglia S., 1992. 
  In: Giampapa M.S., Bookbinder J.A.~(eds.) Proc.~Seventh Cambridge
    Workshop, Cool Stars, Stellar Systems and the Sun. 
    ASP Conf.~Series 26, San Francisco, p.~546

\bibitem{aas} Rutten R.J., 1995. 
  In: Hoeksema J.T., Domingo V., Fleck B., Battrick B.~(eds.) 
    Proc.~4th SOHO Workshop. ESA SP-376, p.~151 
  
\bibitem{aat} Rutten R.J., 1996. 
  In: Strassmeier K.G, Linsky J.L.~(eds.) IAU Symp.~176, 
    Stellar Surface Structure. Kluwer, Dordrecht, 
    p.~385
  
\bibitem{aau} Rutten R.J., Uitenbroek H., 1991, Solar Physics 134, 15
 
\bibitem{aav} Sutmann G., Ulmschneider P., 1995a, A\&A 294, 232 (Paper I)

\bibitem{aaw} Sutmann G., Ulmschneider P., 1995b, A\&A 294, 241 (Paper II)
  
\bibitem{aax} Sutmann G., Musielak Z.E., Ulmschneider P., 1997,
       A\&A (submitted) (Paper III)

\bibitem{aay} Ulmschneider P., 1990. 
  In: Wallerstein G.~(ed.) Proc.~Sixth Cambridge Workshop, Cool Stars,
      Stellar Systems and the Sun.
      ASP Conf.~Series 9, San Francisco, p.~3         
 
\bibitem{aaz} Ulmschneider P., Rammacher W., Gail H.-P., 1992. 
    In: Giampapa M.S., Bookbinder J.A.~(eds.) 
    Proc.~Seventh Cambridge Workshop, Cool Stars, Stellar Systems, and
    the Sun. 
    ASP Conf.~Series 26, San Francisco, p.~471
 
\end{thebibliography}
\end{document}